\documentclass[a4paper,11pt]{article}

\usepackage{amssymb}
\usepackage[T1]{fontenc} 
\usepackage{physics}
\usepackage{comment}
\usepackage{amsmath}
\usepackage{graphicx}
\usepackage{mathtools}
\usepackage{array}
\usepackage{float}
\usepackage{xcolor}
\usepackage{bm}
\usepackage{braket}
\usepackage{cite}

\definecolor{ceruleanblue}{rgb}{0.16, 0.32, 0.75}

\usepackage[linktocpage=true]{hyperref}
\hypersetup{
colorlinks=true,
citecolor=red,
linkcolor=blue,
urlcolor=magenta,
pdfauthor={},
pdftitle={},
pdfsubject={}
}

\numberwithin{equation}{section} 

\makeatletter
\renewcommand\section{\@startsection {section}{1}{\z@}%
                               {-3.5ex \@plus -1ex \@minus -.2ex}
                               {2.3ex \@plus.2ex}%
                               {\normalfont\large\bfseries}}
\renewcommand\subsection{\@startsection{subsection}{2}{\z@}%
                                 {-3.25ex\@plus -1ex \@minus -.2ex}%
                                 {1.5ex \@plus .2ex}%
                                 {\normalfont\bfseries}}
\makeatother

\newfont{\goth}{ygoth.tfm scaled 1200}                   

\begin{document}
\begin{titlepage}
\begin{center}

\hfill         \phantom{xxx}

\begin{flushright}
OCU-PHYS-618
\end{flushright}
\vskip 1.0 cm{\LARGE \bf Vacuum structure of a scalar field}
\vskip 0.3 cm {\LARGE \bf on a torus with uniform magnetic flux}
%
\renewcommand{\thefootnote}{\fnsymbol{footnote}}
\vskip 1.25 cm {\bf Mayumi Akamatsu${}^{1}$, Hiroki Imai${}^{2,3}$
, Makoto Sakamoto${}^{3}$, Maki Takeuchi${}^{4}$
}
\nonumber\\

\vskip 0.2 cm
 {\it ${}^{1}$ Mitsubishi Electric Corporation, Kamakura-shi, Kanagawa, 247-8501, Japan}

\vskip 0.2 cm
 {\it ${}^{2}$ Department of Physics, Graduate School of Science, Osaka Metropolitan University, 3-3-138 Sugimoto, Sumiyoshi, Osaka, 558-8585, Japan}
 
 \vskip 0.2 cm
 {\it ${}^{3}$ Osaka Central Advanced Mathematical Institute (OCAMI), Osaka Metropolitan University, 3-3-138 Sugimoto, Sumiyoshi, Osaka, 558-8585, Japan}

\vskip 0.2 cm
{\it ${}^{4}$ Graduate School of Sciences and Technology for Innovation,
Yamaguchi University, Yamaguchi-shi, Yamaguchi 753-8512, Japan}

\vskip 0.2 cm

\end{center}
\vskip 1.5 cm

%
\begin{abstract}
%
\noindent
\baselineskip=18pt
We investigate the vacuum expectation value of a complex scalar field on a
two-dimensional torus with quantized magnetic flux $M$.
A characteristic feature of this system is the emergence of
a critical area:
when the area of the torus exceeds this critical value, the vacuum expectation
value becomes nonvanishing.
Furthermore, any nonzero vacuum expectation value necessarily exhibits
nontrivial dependence on the coordinates of the torus.
Employing the lowest-mode approximation, we find a single vacuum configuration
for $M=1$, whereas two and six degenerate vacuum configurations arise 
for $M=2$ and $M=3$, respectively.
We then analyze the symmetry properties of these vacuum configurations 
and determine whether they preserve or spontaneously break the symmetry
of the underlying system.
\end{abstract}
\end{titlepage}
\tableofcontents
\flushbottom

%
\section{Introduction}
\label{sec:intro}
If a scalar field coupled to a gauge field acquires a nonzero vacuum expectation value, the gauge symmetry is spontaneously broken and the corresponding gauge bosons become massive \cite{Higgs:1964pj, Higgs:1966ev}. This mechanism, known as the Higgs mechanism, plays a central role in our understanding of the electroweak interaction in the Standard Model of particle physics \cite{Weinberg:1967tq,Salam:1968rm}, as well as in superconductivity in condensed matter physics \cite{Anderson:1963pc}. In the Standard Model, another important role of the vacuum expectation value is to generate masses for the quarks and leptons \cite{Weinberg:1967tq, Salam:1968rm}, which are given as Weyl fermions. Fermion masses arise from three-point interactions among the left-handed fermions, the scalar field, and the right-handed fermions, known as Yukawa couplings. If the scalar field has a constant vacuum expectation value, the Yukawa couplings provide mass terms for fermions in the vacuum. Therefore, the constant vacuum expectation value accounts for the origin of fermion masses. 

In general, the vacuum expectation value of a scalar field could depend on spacetime. However, since the Poincar\'e invariance of spacetime guarantees that it is constant, and most research in particle physics has focused on spacetime-independent vacuum expectation values.
This invariance can be spontaneously broken by dimensional reduction with nontrivial boundary conditions \cite{Sakamoto:1999yk}. In such a situation, the vacuum expectation value can depend on the coordinates of the compactified directions; otherwise, it vanishes. 
This mechanism has been proposed in the context of extra-dimensional models, which are among the attempts to go beyond the Standard Model. Typical analyses with such boundary conditions assume the infinite-volume limit, in which one considers kinks, domain walls, or defects.
In contrast, in systems with extra dimensions of finite volume, the coordinate dependence must be taken into account. An interesting consequence of this coordinate dependence is the existence of a critical size that separates the phases with vanishing and nonvanishing vacuum expectation values \cite{Sakamoto:1999yk,Ohnishi:2000hs,Hatanaka:2000zq,Matsumoto:2001fp,Sakamoto:2001gn,Fujimoto:2011kf}. 
 
A particularly interesting example is provided by field theories on a magnetized torus \cite{Cremades:2004wa,Berasaluce-Gonzalez:2012abm,Blumenhagen:2000wh}. When a constant magnetic flux threads a two-dimensional torus, the fields obey twisted boundary conditions determined by the flux. The presence of magnetic flux leads to a nontrivial degeneracy structure analogous to Landau levels, and the wavefunctions acquire characteristic spatial profiles on the torus. Systems on a magnetized torus have been extensively studied in the context of higher-dimensional field theories and string-motivated models, where they provide a mechanism for generating chiral fermions \cite{Abe:2008sx,Abe:2015yva,Libanov:2000uf,Frere:2000dc,PhysRevD.65.044004,PhysRevD.73.085007,Gogberashvili:2007gg,Guo:2008ia,PhysRevLett.108.181807} and flavor structures\cite{Cremades:2004wa,Arkani-Hamed:1999ylh,Dvali:2000ha,Gherghetta:2000qt,Kaplan:2000av,Huber:2000ie,Fujimoto:2012wv,PhysRevD.97.115039,PhysRevD.90.105006,PhysRevD.88.115007,Buchmuller:2017vho,PhysRevD.97.075019,Kobayashi:2025znw}.
In such a background, it is natural to ask how the vacuum structure of scalar fields is modified. In particular, the interplay between the magnetic flux and the compact geometry may lead to nontrivial vacuum configurations that depend on the torus coordinates. An important question is whether the scalar field can develop a nonvanishing vacuum expectation value and how the vacuum structure depends on the size and moduli of the torus.

In this paper, we investigate the vacuum structure of a complex scalar field on a magnetized two-dimensional torus. We show that the system exhibits a critical value of the torus area that separates two phases. When the torus area is smaller than this critical value, the vacuum expectation value vanishes, whereas when it exceeds the critical value, the scalar field acquires a nonvanishing vacuum expectation value. In contrast to the conventional Higgs mechanism with a constant vacuum expectation value, the vacuum configuration in our system necessarily depends on the torus coordinates due to the presence of the magnetic flux.
We analyze the vacuum configurations using the lowest-mode approximation of the scalar field on the magnetized torus. For the cases with magnetic flux $M=1, 2,$ and $3$, we determine the vacuum configurations that minimize the potential and examine their symmetry properties. We show that the system possesses a discrete symmetry group composed of discrete translations and a rotational symmetry. Depending on the value of the magnetic flux, the vacuum configurations exhibit different patterns of spontaneous symmetry breaking of this discrete symmetry.

 The paper is organized as follows:
 In Section \ref{sec:setup}, we review the setup of a complex scalar field on a magnetized torus. In Section \ref{sec:CA}, we show the existence of a critical area above which the vacuum expectation value becomes nonvanishing.
In Section \ref{sec:VEV}, we analyze the vacuum expectation values for the cases $M=1, 2,$ and 3.
In Section \ref{sec:symmetry}, we analyze the discrete symmetries of the vacuum configurations and clarify whether these configurations preserve or spontaneously break the symmetries of the system. 
Section \ref{Stability} is devoted to the discussion of the stability of the vacuum.
Finally, Section \ref{sec:final} presents our conclusions and discussions.

%
\section{Setup}
\label{sec:setup}
In this paper, we consider a complex scalar field on a two-dimensional torus 
in the presence of a background magnetic flux.
In this section, we present the setup of the system.

\subsection{Magnetized torus}
\begin{figure}[t]
        \centering
        \includegraphics[width=8cm]{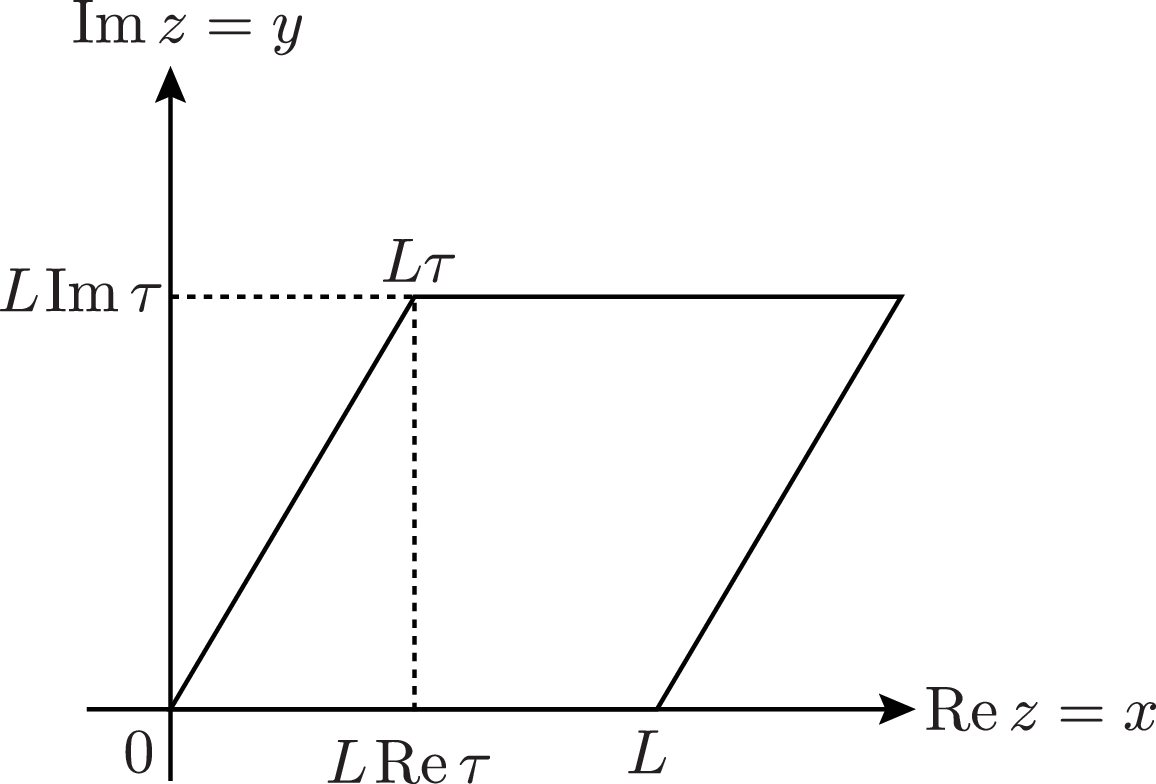}
        \vspace{-6pt}
        \caption{A two-dimensional torus}
        \label{fig_torus}
%
\end{figure}

We define a two-dimensional torus $T^{2}$ as the quotient
space $\mathbb{R}^{2}/\Lambda$, where
$\Lambda\equiv\{n_1\bm{e}_1+n_2\bm{e}_2|n_1,n_2\in\mathbb{Z}\}$
is the lattice in $\mathbb{R}^{2}$ generated by linearly independent 
basis vectors $\bm{e}_i \, (i=1,2)$ given by
%
\begin{align}
    \bm{e}_1=(L,0),\quad \bm{e}_2=(L\,{\rm{Re}}\tau,L\,{\rm{Im}}\tau),
\end{align}
%
where $L$ denotes the overall length scale of $T^2$ (see Figure \ref{fig_torus}).
Two points $\bm{x},\bm{y}\in\mathbb{R}^2$ are identified if 
$\bm{x}-\bm{y}\in\Lambda$, i.e. any $\bm{x}\in T^2$ satisfies
%
\begin{equation}
    \bm{x}\sim \bm{x}+n_1\bm{e}_1+n_2\bm{e}_2
    \label{id_1}
\end{equation}
%
for all $n_1,n_2\in\mathbb{Z}$.

It is convenient to introduce a complex coordinate $Z = x +iy$, 
where $x$ and $y$ are Cartesian coordinates on $T^2$. 
Then, the identification \eqref{id_1} is rewritten as
%
\begin{equation}
    Z\sim Z+n_1 L + n_2 L\tau.
    \label{id_2}
\end{equation}
%
The complex parameter $\tau$, which is called the torus modulus, is taken to 
be $\textrm{Im} \tau > 0$.
The area of the torus $T^{2}$ is given by
%
\begin{equation}
A_{T^2} = L^2 \mathrm{Im}\tau.
    \label{area}
\end{equation}
%
As will be shown in the next section, the area $A_{T^{2}}$ plays a crucial
role in determining the phase structure of the system.
%
%

%
We introduce a background $U(1)$ gauge field $A(Z)$ corresponding to a uniform 
magnetic flux $f$. 
The magnetic flux is given by the 1-form gauge field:
%
\begin{equation}
    A(Z)=\frac{f}{2A_{T^2}}\mathrm{Im}\left\{(\bar{Z}+\bar{a})dZ\right\},
    \label{gauge}
\end{equation}
%
where $a$ is a constant Wilson line phase and 
${\bar{Z}}$ is the complex conjugate of $Z$. 
The gauge field \eqref{gauge} is periodic under the lattice shifts of the torus
(i.e. $Z\mapsto Z+L$ and $Z\mapsto Z+L\tau$)
up to $U(1)$ gauge transformations as
%
\begin{align}
    A(Z+L)&=A(Z)+d\Lambda_1(Z), \label{+L}\\
    A(Z+L\tau)&=A(Z)+d\Lambda_{\tau}(Z), \label{+Ltau}
\end{align}
%
where $\Lambda_1(Z)$ and $\Lambda_{\tau}(Z)$ are gauge parameters given by
%
\begin{align}
    &\Lambda_1(Z)=\frac{f}{2L \,\mathrm{Im}\tau}\mathrm{Im}(Z+a), \label{parameter1}\\
    &\Lambda_{\tau}(Z)=\frac{f}{2L\,\mathrm{Im}\tau}\mathrm{Im}\{\bar\tau(Z+a)\}. \label{parameter2}
\end{align}
%

\subsection{Complex scalar field on $M^d\times T^2$}
In this paper, we consider a complex scalar field $\Phi$ on the spacetime 
$M^d\times T^2$, 
where $M^d$ is the $d$-dimensional Minkowski spacetime and the magnetic flux 
is imposed on $T^2$. 
The Lagrangian is given by
%
\begin{equation}
    \mathcal{L}=-\partial_\mu\Phi^*(x,Z)\partial^\mu\Phi(x,Z)-\sum_{a=d+1}^{d+2}\left|D_{a}\Phi(x,Z)\right|^2 -V(\Phi(x,Z)),
\end{equation}
%
where $\mu$ is the $d$-dimensional spacetime index. 
$D_a=\partial_a -iq A_a$ is the covariant derivative on $T^{2}$
and $q$ denotes the $U(1)$ charge. 
The scalar potential $V(\Phi(x,Z))$ is taken to be of the Higgs-type,
%
\begin{align}
    V(\Phi(x,Z))=-\mu^2|\Phi(x,Z)|^2+\lambda|\Phi(x,Z)|^4 \qquad (\mu^2,\lambda>0).
\label{Higg-like}
\end{align}
%

%
The Hamiltonian reads
%
\begin{equation}
    H=\int d^{d-1} x\int_{T^{2}}\!\!d^2Z \, \bigg\{|\pi(x,Z)|^2+\sum_{j=1}^{d-1}|\partial_j\Phi(x,Z)|^2+|D_{d+1}\Phi(x,Z)|^2+|D_{d+2}\Phi(x,Z)|^2+V(\Phi(x,Z))\bigg\},
    \label{H_scalar_full}
\end{equation}
%
where $\pi(x,Z)\equiv \frac{\partial {\mathcal{L}}}{\partial \dot{\Phi}(x,Z)}$
is the canonical momentum conjugate to $\Phi(x,Z)$.
We are interested in the vacuum expectation value $\langle \Phi(x,Z) \rangle$
that minimizes the Hamiltonian \eqref{H_scalar_full}.
Assuming translational invariance in $M^{d}$, the vacuum expectation
value is independent of $x^{\mu}$, while it may
depend on the coordinate $Z$ on the extra-dimensional torus $T^{2}$.

Focusing on the extra-dimensional dependence, we define
%
\begin{align}
    \widetilde{H}&\equiv \int_{T^2}d^2Z \, \left\{|D_{d+1}\Phi(Z)|^2+|D_{d+2}\Phi(Z)|^2-\mu^2|\Phi(Z)|^2+\lambda|\Phi(Z)|^4\right\} \notag \\
    &=\int_{T^2}d^2 Z \, \left\{-2\Phi^{\ast}(Z)(D_{Z}D_{\bar{Z}}+D_{\bar{Z}}D_{Z})\Phi(Z)-\mu^2|\Phi(Z)|^2+\lambda|\Phi(Z)|^4\right\} ,
    \label{H_scalar}
\end{align}
%
where $D_{Z}$ and $D_{\bar{Z}}$ are covariant derivatives with
respect to the complex coordinate $Z$ and $\bar{Z}$, respectively.
Introducing the dimensionless complex coordinate $z = Z/L$ and the field
$\varphi = \Phi L^{d/2}$, $\widetilde{H}$ can be rewritten as
%
\begin{align}
\widetilde{H}
    &=\frac{1}{L^d}\int_{T^2}d^2 z \, \left\{-2\varphi^{\ast}(z)(D_{z}D_{\bar{z}}
      +D_{\bar{z}}D_{z})\varphi(z)-{\mu^{\prime}}^2|\varphi(z)|^2
      +\lambda^{\prime}|\varphi(z)|^4\right\}
      \notag\\
    &\equiv \frac{1}{L^d}\widetilde{V}[\varphi],
    \label{H_scalar_zerodim}
\end{align}
%
where $\mu^{\prime} = \mu L$ and $\lambda^{\prime} = \lambda L^{-d+2}$
are dimensionless quantities.
The vacuum expectation value $\langle \varphi(z) \rangle$ is determined by
the configuration that minimizes the \lq\lq\,potential \rq\rq\ 
$\widetilde{V}[\varphi]$.
It should be emphasized that $\widetilde{V}[\varphi]$ is a functional
of $\varphi(z)$ but not a function of it.

The scalar field is required to satisfy the pseudo-periodic boundary conditions associated with the gauge transformations:
%
\begin{align}
    \varphi(z+1)&=e^{2\pi i \alpha_1}e^{iq\Lambda_1(z)}\varphi(z),\label{BC1}\\
    \varphi(z+\tau)&=e^{2\pi i \alpha_{\tau}}e^{iq\Lambda_{\tau}(z)}\varphi(z),\label{BC2}
\end{align}
%
where $\alpha_1$ and $\alpha_{\tau}$ are constants called the Scherk-Schwarz phase.

As discussed in Ref.\cite{Abe:2014noa}, the Wilson line phase and 
the Scherk-Schwarz phase are not independent of each other,
so that the Scherk-Schwarz phase can be gauged away without loss of generality.
The Wilson line phase always appears in the combination $Z+a$ 
(see Eq.\eqref{gauge}).
This implies that the Wilson line phase merely plays the role of 
a shift of the origin and does not change the value of the potential 
\eqref{H_scalar_zerodim}.
Therefore, the Wilson line and the Scherk-Schwarz phases are not
important in our analysis and we set them to zero in the following.

As claimed in Refs.\cite{Bachas:1995ik,Abouelsaood:1986gd}, 
the pseudo-periodic boundary conditions \eqref{BC1} and \eqref{BC2}
are well-defined on the torus if and only if the homogeneous flux
is quantized as
%
\begin{align}
    \frac{qf}{2\pi}=M\qquad (M\in {\mathbb{Z}}).
    \label{quantized}
\end{align}
%
To make our analysis simple, we restrict ourselves to $M > 0$,
although one can analyze the case of $M < 0$ in a similar way.

The scalar field $\varphi(z)$ can be always expanded in Kaluza-Klein modes 
on the magnetized torus $T^2$ as
%
\begin{equation}
    \varphi(z)=\sum_{n=0}^\infty \sum_{j=0}^{M-1} a_n^{(j)} \phi_n^{(j,M)}(z),
    \label{mode-expansion}
\end{equation}
%
where $\phi_n^{(j,M)}(z)$ denotes the mode function satisfying 
the following eigenvalue equation:
%
\begin{align}
    -2(D_{z}D_{\bar{z}}+D_{\bar{z}}D_{z})\phi_n^{(j,M)}(z)
    =m_n^2\phi_n^{(j,M)}(z)
\end{align}
%
for $n=0,1,2,\cdots$ and $j=0,1,\cdots,M-1$
with the mass eigenvalue
%
\begin{align}
    m_n^2=\frac{4\pi M}{{\rm{Im}}\tau}\left(n+\frac{1}{2}\right)
    \qquad (n=0,1,2,\cdots).
\end{align}
%
The orthonormal relation is
%
\begin{equation}
   \int_{T^2}\! d^2z \, \phi_n^{(j,M)*}(z)\phi_m^{(k,M)}(z)
    =\delta_{n,m}\delta_{j,k}.
    \label{orthonormal}
\end{equation}
%
These mode functions form an orthonormal basis on $T^{2}$.

The zero mode function $\phi_{0}^{(j,m)}(z)\ (j=0,1,\cdots,M-1)$
is explicitly given by
%
\begin{align}
  \phi_0^{(j,M)}(z)=
  \mathcal{N}
  e^{i\pi M z\frac{{\rm{Im}}z}{{\rm{Im}}\tau}}
   \vartheta
\begin{bmatrix}
\tfrac{j}{M} \\[3pt] 0
\end{bmatrix}
(Mz, M\tau) ,
\end{align}
%
where $ \mathcal{N}$ is a normalization constant determined by 
Eq.\eqref{orthonormal} as
%
\begin{align}
\mathcal{N} = \bigg[ \frac{2M}{\textrm{Im} \tau} \bigg]^{\frac{1}{4}},
\end{align}
%
and the Jacobi theta function is defined by
%
\begin{align}
  \vartheta
\begin{bmatrix}
a \\[3pt] b
\end{bmatrix}
(c, d)=  \sum_{l=-\infty}^{\infty}
e^{i\pi (a+l)^2d}e^{2i \pi (a+l)(c+b)}.
\end{align}
%

%
\section{Critical area}
\label{sec:CA}
%
In this section, we show the existence of a critical area 
above which the vacuum expectation value $\langle \varphi(z) \rangle$
becomes nonvanishing.

The vacuum expectation value $\langle \varphi(z) \rangle$ is 
determined by minimizing the functional $\widetilde{V}[\varphi]$. 
Substituting the mode expansion \eqref{mode-expansion} into the potential \eqref{H_scalar_zerodim}, we obtain
%
\begin{align}
\widetilde{V}[\varphi]
    =\sum_{n=0}^\infty \sum_{j=0}^{M-1}\left\{\frac{4\pi M}{\mathrm{Im}\tau}\left(n+\frac12\right)-{\mu^{\prime}}^2\right\}\left|a_n^{(j)}\right|^2+\lambda^{\prime} \int_{T^2}\! d^2z\, |\varphi(z)|^4,
    \label{momentum-rep}
\end{align}
%
where we have used the orthonormal relation \eqref{orthonormal}.

If all quadratic coefficients 
$\frac{4\pi M}{\mathrm{Im}\tau}\left(n+\frac12\right)-{\mu^{\prime}}^2$
are non-negative, the potential $\widetilde{V}[\varphi]$ is positive-definite
for any $\varphi(z) \ne 0$.
Hence, the minimum of $\widetilde{V}[\varphi]$ is achieved at $\varphi(z)=0$,
implying a vanishing vacuum expectation value.
This occurs when
%
\begin{align}
\frac{2\pi M}{\mathrm{Im}\tau}-{\mu^{\prime}}^2 \ge 0,
    \label{unbroken}
\end{align}
%

\begin{figure}[t]
        \centering
        \includegraphics[width=8cm]{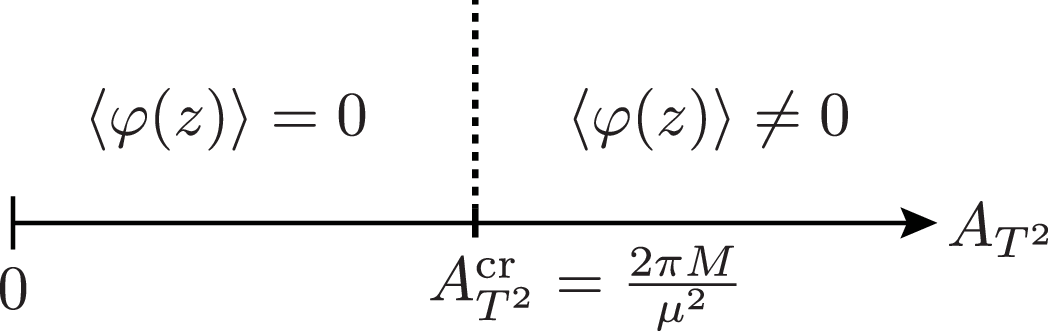}
        \vspace{-6pt}
        \caption{Critical area}
        \label{fig_critical}
%
\end{figure}

%
Conversely, if
%
\begin{align}
\frac{2\pi M}{\mathrm{Im}\tau}-{\mu^{\prime}}^2 < 0,
    \label{symbreaking1}
\end{align}
%
some modes become tachyonic, and the potential is minimized by
a nonvanishing configuration.
The condition \eqref{symbreaking1} can be rewritten as
%
\begin{align}
A_{T^{2}} > A_{T^{2}}^{\mathrm{cr}} \equiv \frac{2\pi M}{\mu^{2}}
   \qquad (A_{T^{2}} = L^{2} \textrm{Im} \tau),
    \label{symbreaking2} 
\end{align}
%
where $A_{T^{2}}^{\mathrm{cr}}$ denotes the critical area above which
the vacuum expectation value $\langle \varphi(z) \rangle$ becomes
nonvanishing (see Figure \ref{fig_critical}).

Thus, we conclude that
%
\begin{align}
\langle \varphi(z) \rangle=0& 
 \qquad {\rm{for}}\quad A_{T^2}\leq A_{T^2}^{\rm{cr}},
   \\
\langle \varphi(z) \rangle\neq 0& 
 \qquad {\rm{for}}\quad A_{T^2}>A_{T^2}^{\rm{cr}}.
\end{align}
%
In the latter case, the $U(1)$ symmetry is spontaneously broken.
%
%

%
A notable feature of this system is that
any nonvanishing vacuum expectation value $\langle \varphi(z) \rangle$
necessarily acquires the coordinate dependence on $T^{2}$, i.e.
%
\begin{align}
\langle \varphi(z) \rangle \ne 0\ \ 
 \longrightarrow\ \ \langle \varphi(z) \rangle\ 
  \textrm{depends on $z$.}
\end{align}
%
This property comes from the observation that a nonzero constant
wavefunction is inconsistent with the boundary conditions
\eqref{BC1} and \eqref{BC2}, and that any nonzero wavefunction
has to depend on the coordinate $z$ to obey the
boundary conditions.

%
\section{Analysis of vacuum expectation values}
\label{sec:VEV}
%
In this section, we analyze the vacuum expectation value
$\langle \varphi(z) \rangle$ for the cases $M = 1, 2,$ and $3$ 
under the condition $A_{T^{2}} > A_{T^{2}}^{\textrm{cr}}$. 
For the purpose of numerical evaluation, we fix the torus moduli 
to $\tau = i$. In this case, the fundamental region of the torus 
corresponds to a square (see Figure~\ref{fig_torus}) and exhibits 
a $\mathbb{Z}_{4}$ rotational symmetry, which will be discussed 
in Section~\ref{sec:symmetry}.

%
\subsection{Lowest-mode approximation of the vacuum}
\label{sub_lowest}
%
 \allowdisplaybreaks
To determine the vacuum expectation value of $\varphi(z)$, 
one must identify the field configuration that minimizes the 
potential $\widetilde{V}[\varphi]$. However, obtaining an exact solution 
is highly nontrivial, because it depends on the coordinate $z$ on $T^{2}$, 
as shown in Section \ref{sec:CA}, and furthermore has to satisfy the nontrivial
boundary conditions \eqref{BC1} and \eqref{BC2}.

To solve the minimization problem approximately, we restrict our
analysis to the parameter region
%
\begin{align}
A_{T^2}^{\rm{cr}} < A_{T^2} < \frac{6\pi M}{\mu^{2}},
\label{area_condition}
\end{align}
%
in which the coefficient $\frac{2\pi M}{\textrm{Im}\tau} - \mu^{\prime2}$
of the quadratic term for $|a^{(j)}_{0}|^{2}$ is negative but other
coefficients for $|a^{(j)}_{n}|^{2}\ (n \ne 0)$ are positive in 
Eq.\eqref{momentum-rep}.
Under this condition, the vacuum expectation value is well
approximated by a linear combination of the lowest-modes
%
\begin{align}
\langle \varphi(z) \rangle 
  \simeq \sum_{j=0}^{M-1} a^{(j)}_{0} \phi^{(j,M)}_{0}(z),
\label{lowest-app}
\end{align}
%
which we refer to as the lowest-mode approximation.
The validity of this approximation will be justified in Section
\ref{Stability}.

Substituting Eq.\eqref{lowest-app} into Eq.\eqref{momentum-rep},
we obtain a potential $\widetilde{V}_{0}$ for the coefficients $a_{0}^{(j)}$:
%
\begin{align}
    \widetilde{V}_0&= \sum_{j=0}^{M-1}\left(\frac{2\pi M}{\mathrm{Im}\tau}-{\mu^{\prime}}^2\right)\left|a_0^{(j)}\right|^2
    +\lambda^{\prime} \int_{T^2}\! d^2z\, |\sum_{j=0}^{M-1} a_0^{(j)} \phi_0^{(j,M)}(z)|^4
    \notag \\
    &=\sum_{j=0}^{M-1}\left(\frac{2\pi M}{\mathrm{Im}\tau}-{\mu^{\prime}}^2\right)\left|a_0^{(j)}\right|^2
    +\lambda^{\prime} \int_{T^2}\! d^2z\, \sum_{j=0}^{M-1} 
    \sum_{j^{\prime}=0}^{M-1}
    \sum_{k=0}^{M-1}
    \sum_{k^{\prime}=0}^{M-1}
    a_0^{(j)\,\ast}  a_0^{(j^{\prime})\,\ast}
     a_0^{(k)}  a_0^{(k^{\prime})}
     \notag \\
     &\qquad \times
      \big( \phi_0^{(j,M)}(z) \phi_0^{(j^{\prime},M)}(z) \big)^{\ast}
      \phi_0^{(k,M)}(z)
     \phi_0^{(k^{\prime},M)}(z)
    \notag \\
    &=\sum_{j=0}^{M-1}\left(\frac{2\pi M}{\mathrm{Im}\tau}-{\mu^{\prime}}^2\right)\left|a_0^{(j)}\right|^2
    +\lambda^{\prime}  \sum_{j=0}^{M-1} 
    \sum_{j^{\prime}=0}^{M-1}
    \sum_{k=0}^{M-1}
    \sum_{k^{\prime}=0}^{M-1}
    \sum_{m=0}^{2M-1}
    \sum_{n=0}^{2M-1}
    a_0^{(j)\,\ast}  a_0^{(j^{\prime})\,\ast}
     a_0^{(k)}  a_0^{(k^{\prime})}
     \notag \\
     &\qquad \times \int_{T^2}\! d^2z\,
       \big(\phi_0^{(j+j^{\prime}+m M,2M)}(z)\big)^{\ast}
        \phi_0^{(k+k^{\prime}+n M,2M)}(z)
      \notag \\
&\qquad    \times  \left(\frac{M}{{\rm{Im}}\tau}\right)^{\frac{1}{2}}
\vartheta
\begin{bmatrix}
\tfrac{j-j^{\prime}+mM}{2M^2} \\[3pt] 0
\end{bmatrix}
(0, -2M^3\bar{\tau})
\,\vartheta
\begin{bmatrix}
\tfrac{k-k^{\prime}+nM}{2M^2} \\[3pt] 0
\end{bmatrix}
(0, 2M^3{\tau})
\notag \\
&=\sum_{j=0}^{M-1}\left(\frac{2\pi M}{\mathrm{Im}\tau}-{\mu^{\prime}}^2\right)\left|a_0^{(j)}\right|^2
    +\lambda^{\prime}  \sum_{j=0}^{M-1} 
    \sum_{j^{\prime}=0}^{M-1}
    \sum_{k=0}^{M-1}
    \sum_{k^{\prime}=0}^{M-1}
    \sum_{m=0}^{2M-1}
    \sum_{n=0}^{2M-1}
    a_0^{(j)\,\ast}  a_0^{(j^{\prime})\,\ast}
     a_0^{(k)}  a_0^{(k^{\prime})}
     \notag \\
     &\qquad \times \delta_{j+j^{\prime}+mM,k+k^{\prime}+nM}^{\text{mod}\,2M}
      \notag \\
&\qquad    \times  \left(\frac{M}{{\rm{Im}}\tau}\right)^{\frac{1}{2}}
\vartheta
\begin{bmatrix}
\tfrac{j-j^{\prime}+mM}{2M^2} \\[3pt] 0
\end{bmatrix}
(0, -2M^3\bar{\tau})
\,\vartheta
\begin{bmatrix}
\tfrac{k-k^{\prime}+nM}{2M^2} \\[3pt] 0
\end{bmatrix}
(0, 2M^3{\tau}),
\label{hamiltonian_zero}
\end{align}
%
where the symbol $\delta_{j,k}^{\text{mod}\,2M}$ is defined by
%
\begin{align}
  \delta_{j,k}^{\text{mod}\,2M}  =\sum_{s \in {\mathbb{Z}}}\delta_{j,k+2sM}.
\end{align}
%
In the third equality of Eq.\eqref{hamiltonian_zero}, we have used
the following theta function identity \cite{Antoniadis:2009bg}
%
\begin{align}
    \vartheta
\begin{bmatrix}
\tfrac{r}{N_1} \\[3pt] 0
\end{bmatrix}
(z_1, \tau N_1)
\cdot
\vartheta
\begin{bmatrix}
\tfrac{s}{N_2} \\[3pt] 0
\end{bmatrix}
(z_2, \tau N_2)
&=\sum_{m \in {\mathbb{Z}}_{N_1+N_2}}
\vartheta
\begin{bmatrix}
\tfrac{r+s+N_1 m}{N_1+N_2} \\[3pt] 0
\end{bmatrix}
(z_1+z_2, \tau (N_1+N_2)) \notag \\
&\quad \times
\vartheta
\begin{bmatrix}
\tfrac{N_2r-N_1s+N_1 N_2 m}{N_1 N_2(N_1+N_2)} \\[3pt] 0
\end{bmatrix}
(z_1 N_2-z_2 N_1, \tau N_1 N_2 (N_1+N_2)),
\end{align}
%
and in the last equality of Eq.\eqref{hamiltonian_zero}, 
we have used the orthonormal relation \eqref{orthonormal}.

%
\subsection{Vacuum configurations for $M=1,2$ and $3$}
\label{subsec:M=1,2,3}
%
%
\subsubsection{$M=1$}
\label{subsubsec:M=1}
%
For $M=1$, the potential \eqref{hamiltonian_zero} reduces to
a function of a single variable $a_{0}^{(0)}$:
%
\begin{align}
    \widetilde{V}_0
    &=\left(\frac{2\pi}{\mathrm{Im}\tau}-{\mu^{\prime}}^2\right)\left|a_0^{(0)}\right|^2+\lambda^{\prime} \left(\frac{1}{{\rm{Im}}\tau}\right)^{\frac{1}{2}}
    |a_0^{(0)}|^4 
    \sum_{m=0}^{1}
    \sum_{n=0}^{1}
\delta_{m,n}^{\text{mod}\,2}\,
    \vartheta
\begin{bmatrix}
\tfrac{m}{2} \\[3pt] 0
\end{bmatrix}
(0, -2\bar{\tau})\,
\vartheta
\begin{bmatrix}
\tfrac{n}{2} \\[3pt] 0
\end{bmatrix}
(0, 2{\tau})
\notag \\
&\equiv  -\alpha \left|a_0^{(0)}\right|^2
+\beta \left|a_0^{(0)}\right|^4,
\label{potential_{M=1}}
\end{align}
%
where the coefficients $\alpha$ and $\beta$ are defined by
%
\begin{align}
\alpha &= -\left(\frac{2\pi}{\mathrm{Im}\tau}-{\mu^{\prime}}^2\right), \\
\beta &= 
    \lambda^{\prime} \left(\frac{1}{{\rm{Im}}\tau}\right)^{\frac{1}{2}}
    \left(\vartheta
\begin{bmatrix}
0 \\[3pt] 0
\end{bmatrix}
(0, -2\bar{\tau})\,
\vartheta
\begin{bmatrix}
0 \\[3pt] 0
\end{bmatrix}
(0, 2{\tau})
+\vartheta
\begin{bmatrix}
\tfrac{1}{2} \\[3pt] 0
\end{bmatrix}
(0, -2\bar{\tau})\,
\vartheta
\begin{bmatrix}
\tfrac{1}{2} \\[3pt] 0
\end{bmatrix}
(0, 2{\tau})
\right).
\end{align} 
%
Since $\alpha$ takes a positive value for the case of Eq.\eqref{area_condition},
the minimum value of the potential \eqref{potential_{M=1}} is realized when
%
\begin{align}
    a_0^{(0)}=\sqrt{\frac{\alpha}{2\beta}},
\end{align}
%
where the phase of $a^{(0)}_{0}$ has been chosen to be real
without loss of generality.
Accordingly, the vacuum configuration for $M=1$ is given by
%
\begin{align}
\langle \varphi^{(M=1)}(z)\rangle
&=\sqrt{\frac{\alpha}{2\beta}}\phi_0^{(0,1)}(z).
\label{VEV_1}
\end{align}
%

%
\subsubsection{$M=2$}
\label{subsubsec:M=2}
%
For $M=2$, Eq.\eqref{hamiltonian_zero} becomes a function of two complex
variables $a_{0}^{(0)}$ and $a_{0}^{(1)}$,
%
\begin{align}
   \widetilde{V}_0&= \sum_{j=0}^{1}\left(\frac{4\pi}{\mathrm{Im}\tau}-{\mu^{\prime}}^2\right)\left|a_0^{(j)}\right|^2
    +\lambda^{\prime}  \sum_{j=0}^{1} 
    \sum_{j^{\prime}=0}^{1}
    \sum_{k=0}^{1}
    \sum_{k^{\prime}=0}^{1}
    \sum_{m=0}^{3}
    \sum_{n=0}^{3}
    a_0^{(j)\,\ast}  a_0^{(j^{\prime})\,\ast}
     a_0^{(k)}  a_0^{(k^{\prime})}
     \notag \\
     &\qquad \times \delta_{j+j^{\prime}+2m,k+k^{\prime}+2n}^{\text{mod}\,4}
      \notag \\
&\qquad    \times  \left(\frac{2}{{\rm{Im}}\tau}\right)^{\frac{1}{2}}
\vartheta
\begin{bmatrix}
\tfrac{j-j^{\prime}+2m}{8} \\[3pt] 0
\end{bmatrix}
(0, -16\bar{\tau})
\,\vartheta
\begin{bmatrix}
\tfrac{k-k^{\prime}+2n}{8} \\[3pt] 0
\end{bmatrix}
(0, 16{\tau})
\notag \\
&\equiv  -\alpha \left(\left|a_0^{(0)}\right|^2+\left|a_0^{(1)}\right|^2\right)
+\beta \left(\left|a_0^{(0)}\right|^4+\left|a_0^{(1)}\right|^4\right)
+\gamma \left|a_0^{(0)}\right|^2\left|a_0^{(1)}\right|^2  \notag \\
&\qquad+\delta \left((a_0^{(0)\ast})^2 (a_0^{(1)})^2+(a_0^{(1)\ast})^2 (a_0^{(0)})^2\right),
\label{potential-M=2}
\end{align}
%
where
%
\begin{align}
\alpha&\equiv -\left(\frac{4\pi}{\mathrm{Im}\tau}-{\mu^{\prime}}^2\right), \\
\beta &\equiv \lambda^{\prime} \left(\frac{2}{{\rm{Im}}\tau}\right)^{\frac{1}{2}}
 \sum_{m=n\,{\text{mod}\,2}}
 \vartheta
\begin{bmatrix}
\tfrac{m}{4} \\[3pt] 0
\end{bmatrix}
(0, -16\bar{\tau})
\,\vartheta
\begin{bmatrix}
\tfrac{n}{4} \\[3pt] 0
\end{bmatrix}
(0, 16{\tau}), \\
\gamma 
  &\equiv 
     \lambda^{\prime} \left(\frac{2}{{\rm{Im}}\tau}\right)^{\frac{1}{2}}
   \sum_{m=n\,{\text{mod}\,2}}
   \Bigg\{
     \vartheta
      \begin{bmatrix}
        \tfrac{-1+2m}{8} \\[3pt] 0
      \end{bmatrix}
    (0, -16\bar{\tau})
    \,\vartheta
      \begin{bmatrix}
        \tfrac{-1+2n}{8} \\[3pt] 0
      \end{bmatrix}
    (0, 16{\tau})
   \notag\\
 &\hspace{5mm}+ 
   \vartheta
     \begin{bmatrix}
       \tfrac{1+2m}{8} \\[3pt] 0
     \end{bmatrix}
    (0, -16\bar{\tau})
    \,\vartheta
      \begin{bmatrix}
        \tfrac{-1+2n}{8} \\[3pt] 0
      \end{bmatrix}
    (0, 16{\tau})
+ 
  \vartheta
    \begin{bmatrix}
      \tfrac{-1+2m}{8} \\[3pt] 0
    \end{bmatrix}
   (0, -16\bar{\tau})
   \,\vartheta
    \begin{bmatrix}
       \tfrac{1+2n}{8} \\[3pt] 0
   \end{bmatrix}
   (0, 16{\tau})
   \notag\\
 &\hspace{5mm}+ 
   \vartheta
    \begin{bmatrix}
      \tfrac{1+2m}{8} \\[3pt] 0
    \end{bmatrix}
   (0, -16\bar{\tau})
   \,\vartheta
    \begin{bmatrix}
      \tfrac{1+2n}{8} \\[3pt] 0
    \end{bmatrix}
   (0, 16{\tau})\Bigg\}, \\
 \delta & \equiv 
 \lambda^{\prime} \left(\frac{2}{{\rm{Im}}\tau}\right)^{\frac{1}{2}}
\sum_{m=n+1\,{\text{mod}\,2}}
 \vartheta
\begin{bmatrix}
\tfrac{m}{4} \\[3pt] 0
\end{bmatrix}
(0, -16\bar{\tau})
\,\vartheta
\begin{bmatrix}
\tfrac{n}{4} \\[3pt] 0
\end{bmatrix}
(0, 16{\tau}).
\end{align}
%
Note that $\alpha > 0$ for the parameter region \eqref{area_condition}.
Since $\delta$ and $2\beta+\gamma-2\delta$ are found to be positive, 
minimizing the potential \eqref{potential-M=2} yields
%
\begin{align}
    a_0^{(0)}&=\sqrt{\frac{\alpha}{2\beta+\gamma-2\delta}},\\
    a_0^{(1)}&=\pm i\sqrt{\frac{\alpha}{2\beta+\gamma-2\delta}},
\end{align}
%
where $a_0^{(0)}$ has been chosen to be real.
Thus, the vacuum configurations for $M=2$ are doubly degenerate and 
given by
%
\begin{align}
\langle \varphi_{\pm}^{(M=2)}(z)\rangle 
&=\sqrt{\frac{\alpha}{2\beta+\gamma-2\delta}}\left(\phi_0^{(0,2)}(z)\pm i\phi_0^{(1,2)}(z)\right).
\label{VEV_2}
\end{align}
%

%
\subsubsection{$M=3$}
\label{subsubsec:M=3}
%
For $M=3$, the potential \eqref{hamiltonian_zero} involves three complex
variables $a_{0}^{(j)}\ (j=0,1,2)$ and takes a more complicated form as
%
\begin{align}
\widetilde{V}_0&=\sum_{j=0}^{2}\left(\frac{6\pi }{\mathrm{Im}\tau}-{\mu^{\prime}}^2\right)\left|a_0^{(j)}\right|^2
    +\lambda^{\prime}  \sum_{j=0}^{2} 
    \sum_{j^{\prime}=0}^{2}
    \sum_{k=0}^{2}
    \sum_{k^{\prime}=0}^{2}
    \sum_{m=0}^{5}
    \sum_{n=0}^5
    a_0^{(j)\,\ast}  a_0^{(j^{\prime})\,\ast}
     a_0^{(k)}  a_0^{(k^{\prime})}
     \notag \\
     &\qquad \times \delta_{j+j^{\prime}+3m,k+k^{\prime}+3n}^{\text{mod}\,6}
        \left(\frac{3}{{\rm{Im}}\tau}\right)^{\frac{1}{2}}
\vartheta
\begin{bmatrix}
\tfrac{j-j^{\prime}+3m}{18} \\[3pt] 0
\end{bmatrix}
(0, -54\bar{\tau})
\,\vartheta
\begin{bmatrix}
\tfrac{k-k^{\prime}+3n}{18} \\[3pt] 0
\end{bmatrix}
(0, 54{\tau}) 
\notag \\
&\equiv  -\alpha \left(\left|a_0^{(0)}\right|^2+\left|a_0^{(1)}\right|^2+\left|a_0^{(2)}\right|^2\right)
+\beta \left(\left|a_0^{(0)}\right|^4+\left|a_0^{(1)}\right|^4+\left|a_0^{(2)}\right|^4\right)
\notag \\
&\qquad
+\gamma \left(\left|a_0^{(0)}\right|^2\left|a_0^{(1)}\right|^2+\left|a_0^{(0)}\right|^2\left|a_0^{(2)}\right|^2+\left|a_0^{(1)}\right|^2\left|a_0^{(2)}\right|^2\right)  \notag \\
&\qquad
+\delta 
  \Big((a_0^{(0)\ast})^2 (a_0^{(1)})(a_0^{(2)})
        + (a_0^{(1)\ast})^2 (a_0^{(0)})(a_0^{(2)})
        + (a_0^{(2)\ast})^2 (a_0^{(0)})(a_0^{(1)})
         \notag\\
&\hspace{10mm}
        + (a_0^{(1)\ast})(a_0^{(2)\ast})(a_0^{(0)})^2
        + (a_0^{(0)\ast})(a_0^{(2)\ast})(a_0^{(1)})^2
        + (a_0^{(0)\ast})(a_0^{(1)\ast})(a_0^{(2)})^2
  \Big),
\label{potential_{M=3}}
\end{align}
%
where
%
\begin{align}
\alpha &=
    -\left(\frac{6\pi }{\mathrm{Im}\tau}-{\mu^{\prime}}^2\right), \\
\beta &=
    \lambda^{\prime} \left(\frac{3}{{\rm{Im}}\tau}\right)^{\frac{1}{2}}
    \sum_{m=n\,{\text{mod}\,2}}
    \vartheta
      \begin{bmatrix}
        \tfrac{m}{6} \\[3pt] 0
      \end{bmatrix}
    (0, -54\bar{\tau})
   \,\vartheta
      \begin{bmatrix}
        \tfrac{n}{6} \\[3pt] 0
      \end{bmatrix}
    (0, 54{\tau}), \\
\gamma &= 
    \lambda^{\prime} \left(\frac{3}{{\rm{Im}}\tau}\right)^{\frac{1}{2}}
     \sum_{m=n\,{\text{mod}\,2}}
      \Bigg\{
        \vartheta
         \begin{bmatrix}
            \tfrac{-1+3m}{18} \\[3pt] 0
         \end{bmatrix}
       (0, -54\bar{\tau})
       \,\vartheta
        \begin{bmatrix}
          \tfrac{-1+3n}{18} \\[3pt] 0
        \end{bmatrix}
       (0, 54{\tau})
       \notag\\
&\hspace{5mm}    +
     \vartheta
      \begin{bmatrix}
         \tfrac{1+3m}{18} \\[3pt] 0
      \end{bmatrix}
     (0, -54\bar{\tau})
     \,\vartheta
      \begin{bmatrix}
          \tfrac{-1+3n}{18} \\[3pt] 0
      \end{bmatrix}
     (0, 54{\tau})
  +
    \vartheta
     \begin{bmatrix}
       \tfrac{-1+3m}{18} \\[3pt] 0
     \end{bmatrix}
    (0, -54\bar{\tau})
    \,\vartheta
     \begin{bmatrix}
       \tfrac{1+3n}{18} \\[3pt] 0
     \end{bmatrix}
    (0, 54{\tau})
    \notag\\
&\hspace{5mm}
  +
   \vartheta
    \begin{bmatrix}
      \tfrac{1+3m}{18} \\[3pt] 0
    \end{bmatrix}
   (0, -54\bar{\tau})
   \,\vartheta
    \begin{bmatrix}
       \tfrac{1+3n}{18} \\[3pt] 0
    \end{bmatrix}
   (0, 54{\tau})\Bigg\}, \\
\delta &=
    \lambda^{\prime} \left(\frac{3}{{\rm{Im}}\tau}\right)^{\frac{1}{2}}
   \sum_{m=n\,{\text{mod}\,2}}
   \Bigg\{
          \vartheta
           \begin{bmatrix}
             \tfrac{m}{6} \\[3pt] 0
           \end{bmatrix}
         (0, -54\bar{\tau})
         \,\vartheta
          \begin{bmatrix}
             \tfrac{-2+3n}{18} \\[3pt] 0
          \end{bmatrix}
         (0, 54{\tau}) 
         \notag\\
&\hspace{5mm}
  +
   \vartheta
    \begin{bmatrix}
       \tfrac{m}{6} \\[3pt] 0
    \end{bmatrix}
   (0, -54\bar{\tau})
   \,\vartheta
    \begin{bmatrix}
       \tfrac{2+3n}{18} \\[3pt] 0
    \end{bmatrix}
    (0, 54{\tau}) \Bigg\}.
\label{potential-M=3}
\end{align}
%
Due to the complexity of the potential,
it seems to be difficult to treat the minimization of the potential
\eqref{potential_{M=3}} analytically, so we numerically search 
the vacuum configuration
%
\begin{align}
\langle \varphi^{(M=3)}
(z) \rangle
&=a\phi_0^{(0,3)}(z)+b\phi_0^{(1,3)}(z)+c\phi_0^{(2,3)}(z).
\label{VEV_M3}
\end{align}
%
Performing a numerical minimization of the potential, 
we find six degenerate vacuum solutions 
for $M=3$, as summarized in Table \ref{Tab:M_3coeff}.
%
\begin{table}[htbp]
  \centering
  \caption{The vacuum solutions for $M=3$ and $\tau=i$}
  \begin{tabular}{cccc}
  \hline
  type & $a$ & $b$ & $c$ \\ \hline
  I &  $-0.0111561+0.0862216i$ & $0.0802481-0.0334494i$ 
  &$0.0802481-0.0334494i$ \\ \hline
  II & $0.0802481-0.0334494i$ &
 $ -0.0111561+0.0862216i$ &
  $0.0802481-0.0334494i$ \\ \hline
  III & $0.0802481-0.0334494i$ &
 $ 0.0802481-0.0334494i$ &
 $ -0.0111561+0.0862216i$ \\ \hline
  IV & $-0.0160683-0.0854426i$ &
  $0.0820296+0.0288056i$ &
  $0.0820296+0.0288056i$  \\ \hline
  V & $0.0820296+0.0288056i$
  & $0.0820296+0.0288056i$ &
 $ -0.0160683-0.0854426i$
  \\ \hline
  VI & $0.0820296+0.0288056i$ &
  $-0.0160683-0.0854426i$ &
  $0.0820296+0.0288056i$ \\ \hline
\end{tabular}
   \label{Tab:M_3coeff}
\end{table}
%

%
\section{Discrete symmetry of the system and the vacuum configurations}
\label{sec:symmetry}
%
In this section, we analyze the discrete symmetries of the vacuum configurations
obtained in Section \ref{sec:VEV} and clarify whether these configurations 
preserve or spontaneously break the symmetries of the system.

%
\subsection{Discrete symmetry of the system}
%
As discussed in Refs.\cite{10.1063/1.1616203, Abe:2014noa}, 
the continuous translational symmetry 
on the torus with quantized magnetic flux $M \in {\mathbb{Z}}\ (M>0)$ 
is broken to a discrete translational one:
%
\begin{align}
\mathbb{Z}_M^1 \times \mathbb{Z}_M^{\tau}.
\end{align}
%
These discrete symmetries correspond to 
the translations of the complex coordinate $z$ on the torus:
%
\begin{align}
\mathbb{Z}_M^{1}&=\{ (t_{1})^n,\,n=0,1,\cdots,M-1\,({\rm{mod}} \,M)\,|\,
  (t_{1})^M=1\}, \\
\mathbb{Z}_M^{\tau}&=\{ (t_{\tau})^n,\,n=0,1,\cdots,M-1\,({\rm{mod}} \,M)\,|\,  
  (t_{\tau})^M=1\},
\end{align}
%
where $t_{1}$ and $t_{\tau}$ are defined by the actions on the complex 
coordinate $z$ as
%
\begin{align}
 t_1(z)&=z-\frac{1}{M}, \label{t_1}\\
t_{\tau}(z)&=z-\frac{\tau}{M}.
\label{t_tau}
\end{align}
%

%
As was expected from Figure \ref{fig_torus}, the torus model with $\tau=i$
possesses an additional $\mathbb{Z}_{4}$ rotational symmetry:
%
\begin{align}
{\mathbb{Z}}_4^{\omega}
  &=\{ (r_{\omega})^m,\,m=0,1,2,3\,({\rm{mod}} \,4)\,|\,(r_{\omega})^{4}=1\},
\end{align}
%
where $r_{\omega}$ acts on the complex coordinate $z$ as
%
\begin{align}
r_{\omega}(z) = \omega z \qquad (\omega=e^{2\pi i/4}=i).
\end{align}
%

%
As discussed in \cite{Abe:2014noa}, the index $j$, which distinguishes 
the degenerate eigenfunctions $\phi_{0}^{(j,M)}(z)\ (j=0,1,\cdots,M-1)$,
is associated with the eigenvalue of the conserved operator 
$e^{2\pi i\hat{\widetilde{Y}}}$, i.e.
%
\begin{align}
e^{2\pi i\hat{\widetilde{Y}}}\phi_0^{(j,M)}(z)
 &=e^{2\pi i\frac{j}{M}}\phi_0^{(j,M)}(z),
\label{Y_trans}
\end{align}
%
where the explicit form of $e^{2\pi i\hat{\widetilde{Y}}}$ 
in the complex coordinate is given by
%
\begin{align}
   e^{2\pi i\hat{\widetilde{Y}}}
    =e^{\frac{i}{M}\left(-i(\partial_z+\partial_{\bar{z}})-\frac{\pi M}{{\rm{Im}}\tau}{\rm{Im}}z\right)}.
    \label{Y}
\end{align}
%
Since the differential operator $e^{\frac{1}{ M}(\partial_z+\partial_{\bar{z}})}$ 
on the right-hand side of Eq.\eqref{Y} is nothing but the translation operator 
by $1/M$ in the ${\rm{Re}}\,z$-direction, 
Eq.\eqref{Y_trans} can be rewritten as
%
\begin{align}
    e^{-\frac{i\pi }{{\rm{Im}}\tau}{\rm{Im}}z}\phi_0^{(j,M)}(z+\tfrac{1}{M})=e^{2\pi i\frac{j}{M}}\phi_0^{(j,M)}(z).
\label{1-translation}
\end{align}
%
Since the phase $e^{-\frac{i\pi }{{\rm{Im}}\tau}{\rm{Im}}z}$
on the left-hand side of Eq.\eqref{1-translation} is originated from the
pseudo-periodic boundary condition \eqref{BC1}, this phase is 
irrelevant in our discussions, so that it is appropriate to rewrite
Eq.\eqref{1-translation} into the following equivalence relation:
%
\begin{align}
    \phi_0^{(j,M)}(z+\tfrac{1}{M})
    \sim
    e^{2\pi i\frac{j}{M}}\phi_0^{(j,M)}(z).
    \label{t_1act}
\end{align}
%

%
We then introduce the operation $T_{1}$, which is associated with $t_{1}$,
acting on the zero mode eigenfunctions $\phi_{0}^{(j,M)}(z)$ as
%
\begin{align}
  T_{1}(\phi_0^{(j,M)})(z) \equiv   e^{2\pi i\frac{j}{M}}\phi_0^{(j,M)}(z).
\label{T_1act}
\end{align}
%
In terms of $t_{1}$ and $T_{1}$, the relation \eqref{t_1act} can 
be written as
%
\begin{align}
\phi_0^{(j,M)}(t_1^{-1}(z))\sim T_{1}(\phi_0^{(j,M)})(z).
\label{T_1}
\end{align}
%

%
Similarly, we have another conserved operator $e^{-\frac{i}{M}\hat{\widetilde{P}}}$
which acts on the zero mode eigenfunctions as a shift in the
degenerate index \cite{Abe:2014noa}:
%
\begin{align}
    e^{-\frac{i}{M}\hat{\widetilde{P}}}
    \phi_0^{(j,M)}(z)
&=\phi_0^{(j+1,M)}(z),
\label{Ptrans}
\end{align}
%
where the explicit form of the operator $e^{-\frac{i}{M}\hat{\widetilde{P}}}$
in the complex coordinate is
%
\begin{align}
    e^{-\frac{i}{M}\hat{\widetilde{P}}}
    =e^{-\frac{i}{M}\left(i(\tau\partial_z+\bar{\tau}\partial_{\bar{z}})-\frac{\pi M}{{\rm{Im}}\tau}{\rm{Im}}[\tau\bar{z}]\right)}.
    \label{P}
\end{align}
%
Since the differential operator 
$e^{\frac{1}{M}(\tau\partial_z+\bar{\tau}\partial_{\bar{z}})}$
on the right-hand side of Eq.\eqref{P} is the translation operator by
$\tau/M$ in the $\tau$-direction, Eq.\eqref{Ptrans} can be rewritten as
%
\begin{align}
    e^{\frac{i\pi }{{\rm{Im}}\tau}{\rm{Im}}[\tau\bar{z}]}\phi_0^{(j,M)}(z+\tfrac{\tau}{M})
&=\phi_0^{(j+1,M)}(z).
\label{P_re}
\end{align}
%
Since the phase factor 
$e^{\frac{i\pi }{{\rm{Im}}\tau}{\rm{Im}}[\tau\bar{z}]}$
on the left-hand side is irrelevant in our discussions, it turns out to be
useful to express Eq.\eqref{P_re} into the following equivalence
relation:
%
\begin{align}
   \phi_0^{(j,M)}(z+\tfrac{\tau}{M})
&\sim \phi_0^{(j+1,M)}(z) .
\label{P_sim}
\end{align}
%

%
It is then convenient to introduce the operator $T_{\tau}$, which is associated
with $t_{\tau}$, acting on the zero mode eigenfunctions as
%
\begin{align}
   T_{\tau}( \phi_0^{(j,M)})(z)\equiv 
   \phi_0^{(j+1,M)}(z).
   \label{T_tau}
\end{align}
%
In terms of $t_{\tau}$ and $T_{\tau}$, the relation \eqref{P_sim}
can be written as
%
\begin{align}
    \phi_0^{(j,M)}(t_{\tau}^{-1}(z))
&\sim  T_{\tau}( \phi_0^{(j,M)})(z).
\end{align}
%

%
We finally discuss the rotational symmetry of the system.
As was expected from Figure \ref{fig_torus} with $\tau=i$,
the system possesses the $\mathbb{Z}_{4}$-rotational symmetry.
Since $\phi_0^{(j,M)}(\omega z)$ with $\omega=e^{2\pi i/4}=i$
turns out to obey the same eigenvalue equation and the boundary
conditions as $\phi_0^{(k,M)}(z)$, $\phi_0^{(j,M)}(\omega z)$
has to be expanded in some linear combination of $\phi_0^{(k,M)}(z)$ as
%
\begin{align}
    \phi_0^{(j,M)}(\omega z)
&=\sum_{k=0}^{M-1}
D_{jk}\,
\phi_0^{(k,M)}(z),
\label{rot}
\end{align}
%
where $\tau=\omega=i$.
The coefficients $D_{jk}$ have been derived in Ref.\cite{Abe:2014noa} as
%
\begin{align}
D_{jk}& = \frac1{\sqrt M} \, e^{2 \pi i \frac{jk}M }.
\label{D_{jk}}
\end{align}
%

%
For later convenience, we introduce the operations $r_{\omega}$
and $R_{\omega}$ whose actions on the complex coordinate $z$
and the zero mode eigenfunctions $\phi_{0}^{(j,k)}(z)$, respectively, 
are defined by
%
\begin{align}
r_{\omega}(z)&=\omega z \qquad (\omega=i),\label{r_omega}\\
R_{\omega}(\phi_0^{(j,M)})(z)
  &= \sum_{k=0}^{M-1} (D_{jk})^{\ast}\,\phi_0^{(k,M)}(z).
\label{R_omega}
\end{align}
%
Thus, the relation \eqref{rot} can be rewritten as
%
\begin{align}
    \phi_0^{(j,M)}(r_{\omega}^{-1}(z) )
&=R_{\omega}(\phi_0^{(j,M)})(z).
\end{align}
%

%
It follows from the above discussions that the discrete symmetry group
$G$ of the system is found to be of the form
%
\begin{align}
    G=(\mathbb{Z}_M^1 \times \mathbb{Z}_M^{\tau} )\rtimes {\mathbb{Z}}_4^{\omega},
    \label{G}
\end{align}
%
where $\rtimes$ denotes a semi-direct product, and the elements of $G$ 
are given by
%
\begin{align}
G&=\{ (t_{1})^n(t_{\tau})^m(r_{\omega})^l,\,n,m=0,1,\cdots,M-1\,({\rm{mod}} \,M), 
       \,l=0,1,2,3\,({\rm{mod}} \,4)|
       \notag\\
&\hspace{5mm}    
       (t_{1})^M=(t_{\tau})^M=1, (r_{\omega})^4=1\}.
\end{align}
%

%
\subsection{Discrete symmetry of the vacuum configurations}
%
In this subsection, we examine the symmetry properties of the vacuum configurations
obtained in Section \ref{sec:VEV} for $M=1,2$ and $3$.

%
\subsubsection{$M=1$}
\label{subsec:M_1}
\begin{figure}[t]
        \centering
        \includegraphics[width=0.3\textwidth]{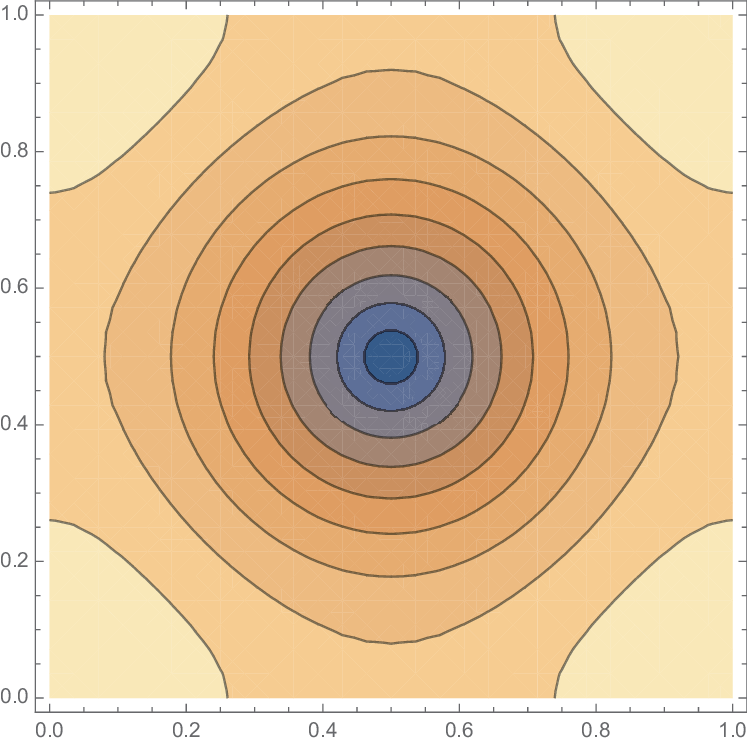} 
        \vspace{-6pt}
        \caption{The zero point of the vacuum solution $\langle \varphi^{(M=1)}(z)\rangle $}
        \label{fig_M_1}
%
\end{figure}

%
Let us first discuss the symmetry of the vacuum configuration for $M=1$. 
From Eq.\eqref{VEV_1}, the vacuum configuration is given by
%
\begin{align}
\langle \varphi^{(M=1)}(z)\rangle
&=\sqrt{\frac{\alpha}{2\beta}}\phi_0^{(0,1)}(z).
\label{VEV_1_2}
\end{align}
%
This configuration has a single zero determined by
%
\begin{align}
   \vartheta
\begin{bmatrix}
0 \\[3pt] 0
\end{bmatrix}
(z, \tau)  
=0,
\end{align}
%
which is located at
%
\begin{align}
z=\frac{1}{2}+\frac{1}{2}\tau,
\end{align}
%
as depicted in Figure \ref{fig_M_1}.
%
%

%
As shown in Eq.\eqref{G}, the system with $M=1$ and $\tau=i$ has the 
discrete symmetry $\mathbb{Z}_{4}^{\omega}$.
(Note that $\mathbb{Z}_{M}^{1} \times \mathbb{Z}_{M}^{\tau}$ is trivial
when $M=1$.)
From Eqs.\eqref{rot} and \eqref{D_{jk}}, we find
%
\begin{align}
\langle \varphi^{(M=1)}(\omega z)\rangle
 &=\langle \varphi^{(M=1)}(z)\rangle 
    \qquad (\omega=i).
\end{align}
%
Therefore, it is concluded that the vacuum configuration \eqref{VEV_1_2}
is invariant under the $\mathbb{Z}_{4}^{\omega}$ rotation
and that the vacuum preserves the full discrete symmetry of the system
for $M=1$ and $\tau=i$.

%
\subsubsection{$M=2$}
\label{subsec:M_2}
%
From Eq.\eqref{G}, the discrete symmetry group $G$ of the system 
for $M=2$ is
%
\begin{align}
G&= (\mathbb{Z}_{M=2}^1 \times \mathbb{Z}_{M=2}^{\tau} )
     \rtimes {\mathbb{Z}}_4^{\omega}
      \notag\\
 &= \{ (t_{1})^n(t_{\tau})^m(r_{\omega})^l,\,n,m=0,1\,({\rm{mod}} \,2),
    \,l=0,1,2,3\,({\rm{mod}} \,4)|\,(t_{1})^2=(t_{\tau})^2=1, (r_{\omega})^4=1\}.
    \label{G_{M=2}}
\end{align}
%
We investigate whether the vacuum configurations \eqref{VEV_2}
for $M=2$ and $\tau=i$ preserve the symmetry $G$ of the 
system or not.

\begin{figure}[t]
\centering
\begin{minipage}{0.49\textwidth}
\centering
\includegraphics[width=0.6\textwidth]{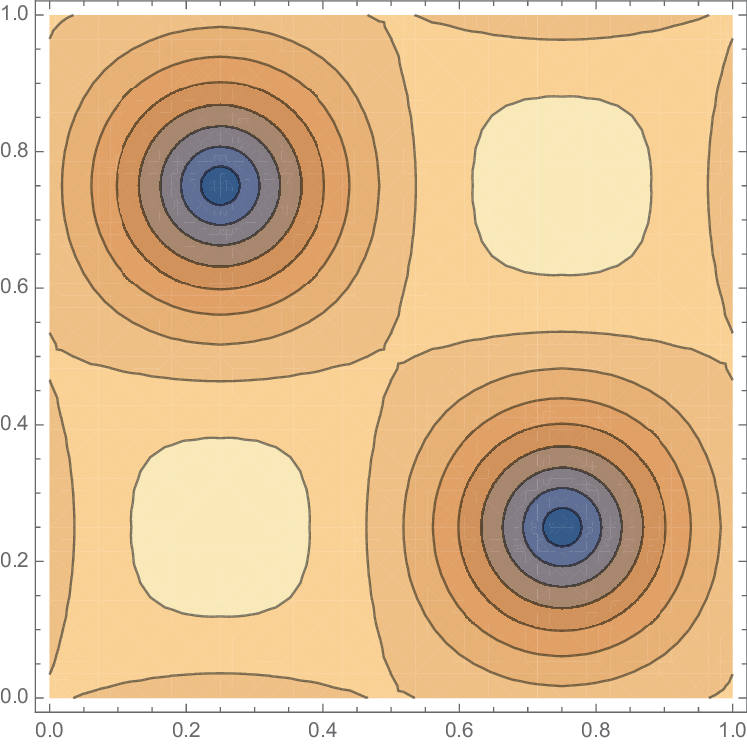} \\
{\small (a) $\langle \varphi^{(M=2)}_{+}\rangle$}
\end{minipage}
\hfill
\begin{minipage}{0.49\textwidth}
\centering
\includegraphics[width=0.6\textwidth]{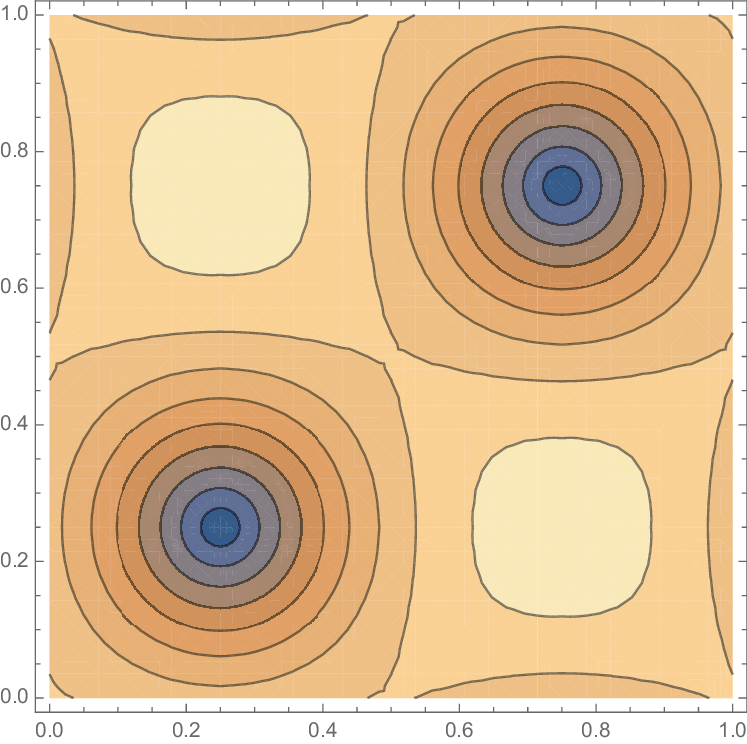} \\
{\small (b) $\langle \varphi^{(M=2)}_{-}\rangle$}
\end{minipage}
\caption{The zero points of the vacuum solution $\varphi^{(M=2)}_{\pm}(z)$}
\label{fig_M_2}
\end{figure}

From Eq.\eqref{VEV_2}, we found two degenerate vacuum configurations
%
\begin{align}
\varphi^{({\rm{I}})}_{\rm{vac}}(z)
&\equiv \langle \varphi^{(M=2)}_{+}(z)\rangle =\sqrt{\frac{\alpha}{2\beta+\gamma-2\delta}}\left(\phi_0^{(0,2)}(z)+i\phi_0^{(1,2)}(z)\right),\\
\varphi^{({\rm{II}})}_{\rm{vac}}(z)
&\equiv \langle \varphi^{(M=2)}_{-}(z)\rangle
=\sqrt{\frac{\alpha}{2\beta+\gamma-2\delta}}\left(\phi_0^{(0,2)}(z)-i\phi_0^{(1,2)}(z)\right).
\label{VEV_M2_00}
\end{align}
%
Each vacuum configuration has two zeros, as shown in Figure \ref{fig_M_2}.
The positions of the zeros are located at
%
\begin{align}
z_1^{\rm{(I)}}=\frac{1+3\tau}{4},\,\,
z_2^{\rm{(I)}}=\frac{3+\tau}{4}
\qquad{\rm{for}}\quad \varphi^{({\rm{I}})}_{\rm{vac}}(z), \\
z_1^{\rm{(II)}}=\frac{1+\tau}{4},\,\,
z_2^{\rm{(II)}}=\frac{3+3\tau}{4}
\qquad{\rm{for}}\quad \varphi^{({\rm{II}})}_{\rm{vac}}(z).
\end{align}
%
%

%
The set of $\{z_1^{\rm{(I)}},z_2^{\rm{(I)}}\}$ 
(and also $\{z_1^{\rm{(II)}},z_2^{\rm{(II)}}\}$)
is found to be invariant under the following transformations:
%
\begin{align}
    H&=\mathbb{Z}_2^{\prime} \times \mathbb{Z}_4^{\prime} 
    \notag \\
    &=\{ (t_{1}t_{\tau})^n(t_1 r_{\omega})^m,\,n=0,1\,({\rm{mod}} \,2),\,m=0,1,2,3\,({\rm{mod}} \,4)\},
\label{H_{M=2}}
\end{align}
%
which is a subgroup of $G$, generated by combinations of translations and
rotations.
We can further show that from Eqs.\eqref{T_1act}, \eqref{T_tau}, and 
\eqref{R_omega} the vacuum configurations 
$\varphi^{({\rm{I,II}})}_{\rm{vac}}(z)$ are invariant under the
transformations $T_{1}T_{\tau}$ and $T_{1}R_{\omega}$, which are associated with
$t_{1}t_{\tau}$ and $t_{1}r_{\omega}$ in Eq.\eqref{H_{M=2}}, such as
%
\begin{align}
    T_1 T_{\tau}(\varphi^{({\rm{I,II}})}_{\rm{vac}}(z)) &\sim (\varphi^{({\rm{I,II}})}_{\rm{vac}}(z)),\\
    T_1 R_{\omega}(\varphi^{({\rm{I,II}})}_{\rm{vac}}(z)) &\sim (\varphi^{({\rm{I,II}})}_{\rm{vac}}(z)).
\end{align}
%
Therefore, we conclude that the symmetry $G$ of the system for $M=2$
and $\tau=i$ is spontaneously broken to $H$, i.e.
%
\begin{align}
G=(\mathbb{Z}_{M=2}^1 \times \mathbb{Z}_{M=2}^{\tau} )
   \rtimes {\mathbb{Z}}_4^{\omega}
    \ \ \longrightarrow\ \ 
     H = \mathbb{Z}_{2}^{\prime} \times \mathbb{Z}_{4}^{\prime}.
\label{symbreaking_{M=2}}
\end{align}
%

%
The above fact of Eq.\eqref{symbreaking_{M=2}} explains the reason
why there appear two degenerate vacuum configurations 
$\varphi^{(\textrm{I})}_{\textrm{vac}}(z)$ and 
$\varphi^{(\textrm{II})}_{\textrm{vac}}(z)$.
Under the transformation of $t_{1}$, which belongs to $G/H$, the set
of $\{ z_{1}^{(\textrm{I})}, z_{2}^{(\textrm{I})} \}$ 
(and also $\{ z_{1}^{(\textrm{II})}, z_{2}^{(\textrm{II})}$ \})
is not invariant but $\{ z_{1}^{(\textrm{I})}, z_{2}^{(\textrm{I})} \}$ 
maps into $\{ z_{1}^{(\textrm{II})}, z_{2}^{(\textrm{II})} \}$, and vice versa.
Correspondingly, the vacuum configuration $\varphi^{(\textrm{I})}_{\textrm{vac}}(z)$
($\varphi^{(\textrm{II})}_{\textrm{vac}}(z)$) maps into 
$\varphi^{(\textrm{II})}_{\textrm{vac}}(z)$ 
($\varphi^{(\textrm{I})}_{\textrm{vac}}(z)$)
by the transformation $T_{1}$, i.e.
%
\begin{align}
\varphi^{(\textrm{I})}_{\textrm{vac}}(z)
 \ \ \stackrel{T_{1}}{\longleftrightarrow}\ \ 
  \varphi^{(\textrm{II})}_{\textrm{vac}}(z).
\end{align}
%
Thus, the two vacuum configurations $\varphi^{(\textrm{I})}_{\textrm{vac}}(z)$ and 
$\varphi^{(\textrm{II})}_{\textrm{vac}}(z)$ are connected each other by the
broken symmetry transformations belonging to the coset $G/H$, 
and hence they are physically equivalent.

%
\subsubsection{$M=3$}
\label{subsec:M_3}
%
We finally analyze the symmetry properties of the vacuum configurations
for $M=3$ and $\tau=i$. 
From Eq.\eqref{G}, the discrete symmetry group $G$ of the system is given by
%
\begin{align}
G&= (\mathbb{Z}_{M=3}^1 \times \mathbb{Z}_{M=3}^{\tau} )
     \rtimes {\mathbb{Z}}_4^{\omega}
      \notag\\
 &= \{ (t_{1})^n(t_{\tau})^m(r_{\omega})^l,\,n,m=0,1,2\,({\rm{mod}} \,3),
    \,l=0,1,2,3\,({\rm{mod}} \,4)|\,(t_{1})^3=(t_{\tau})^3=1, (r_{\omega})^4=1\}.
    \label{G_{M=3}}
\end{align}
%

%
From Eq.\eqref{VEV_M3}, the vacuum solution is parameterized as
%
\begin{align}
\langle \varphi^{(M=3)}(z) \rangle
  =a\phi_0^{(0,3)}(z)+b\phi_0^{(1,3)}(z)+c\phi_0^{(2,3)}(z).
\label{sol_{M=3}}
\end{align}
%
Then, we numerically found the six degenerate vacuum configurations to minimize
the potential, denoted by 
$\varphi^{({\rm{I}})}_{\rm{vac}}(z),\varphi^{({\rm{II}})}_{\rm{vac}}(z),\cdots,\varphi^{({\rm{VI}})}_{\rm{vac}}(z)$.
Their coefficients in 
Eq.\eqref{sol_{M=3}} are summarized in Table \ref{Tab:M_3coeff}.
In the following, we analyze each vacuum configuration individually.

%
\begin{figure}[t]
\centering
\begin{minipage}{0.32\textwidth}
\centering
\includegraphics[width=0.55\textwidth]{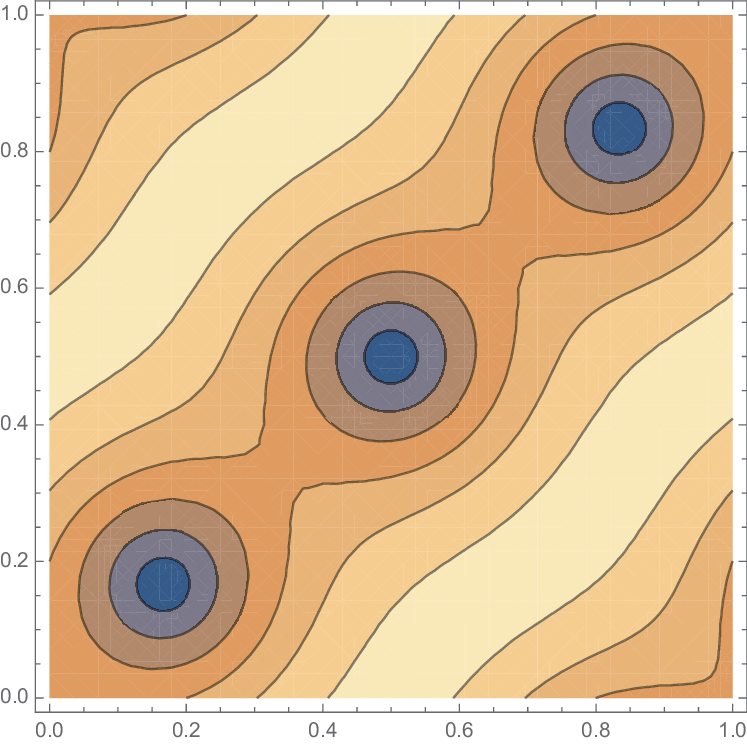} \\
{\small (a) $\langle \varphi^{(M=3)}_{{\rm{I}}}\rangle$}
\end{minipage}
\hfill
\begin{minipage}{0.32\textwidth}
\centering
\includegraphics[width=0.55\textwidth]{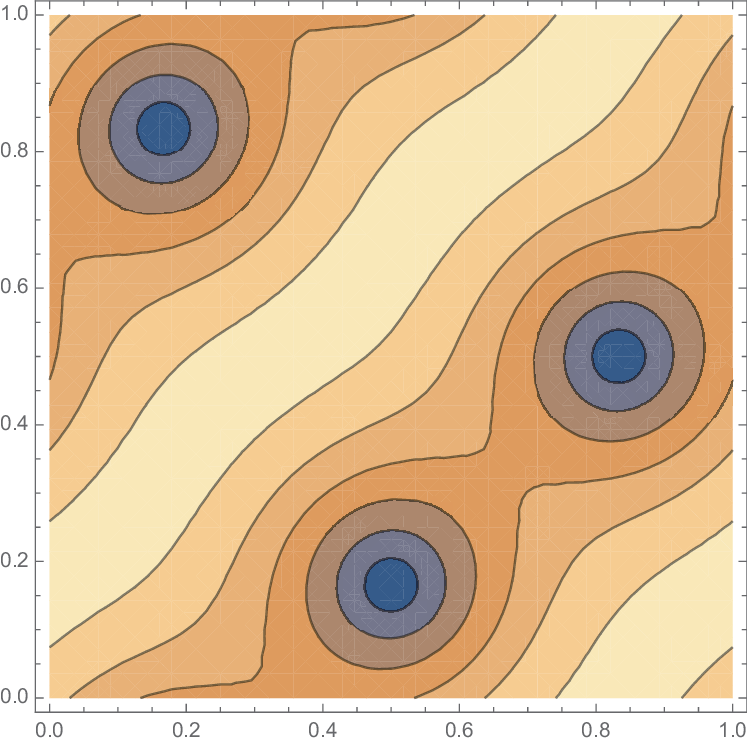} \\
{\small (b) $\langle \varphi^{(M=3)}_{{\rm{II}}}\rangle$}
\end{minipage}
\hfill
\begin{minipage}{0.32\textwidth}
\centering
\includegraphics[width=0.55\textwidth]{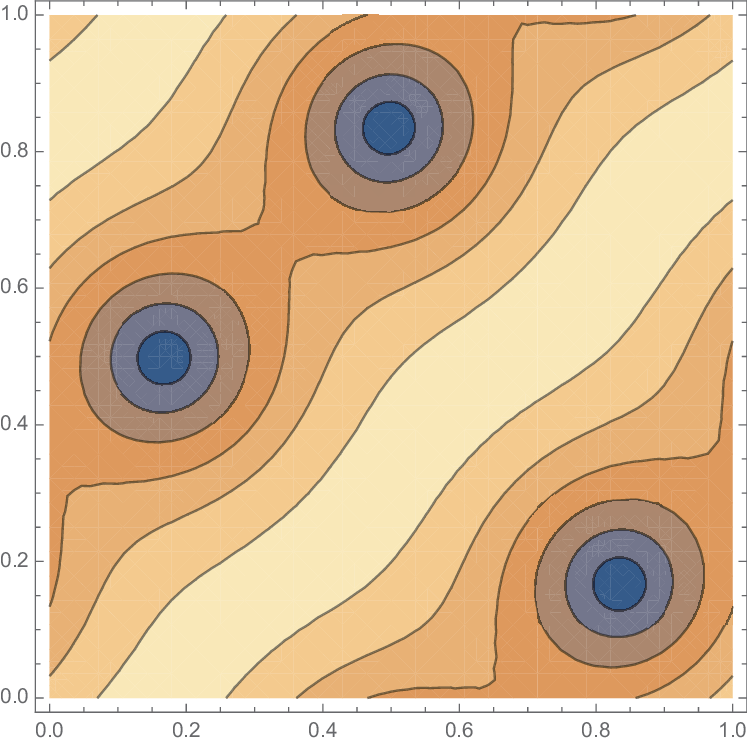} \\
{\small (c) $\langle \varphi^{(M=3)}_{{\rm{III}}}\rangle$}
\end{minipage}
\vspace{3mm}

\begin{minipage}{0.32\textwidth}
\centering
\includegraphics[width=0.55\textwidth]{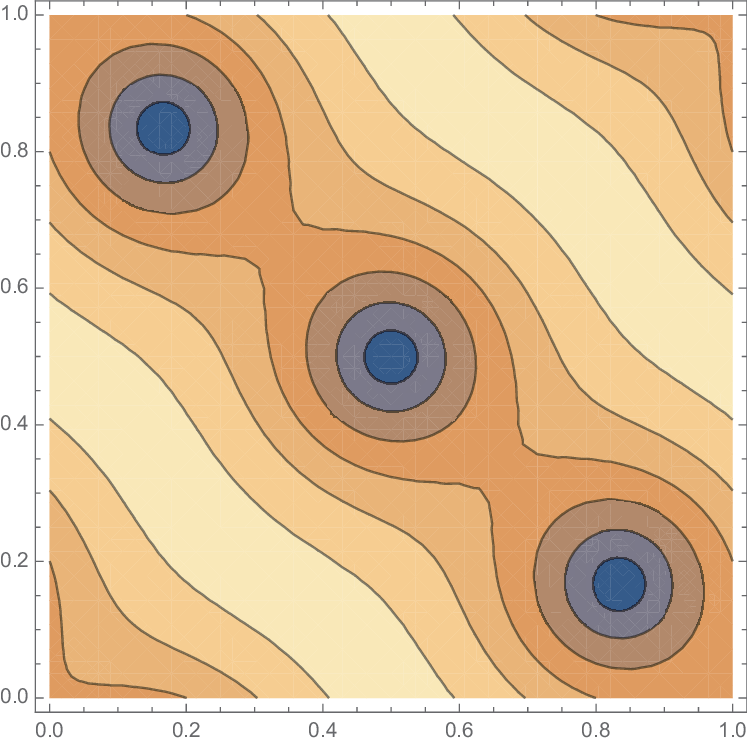} \\
{\small (d) $\langle \varphi^{(M=3)}_{{\rm{IV}}}\rangle$}
\end{minipage}
\hfill
\begin{minipage}{0.32\textwidth}
\centering
\includegraphics[width=0.55\textwidth]{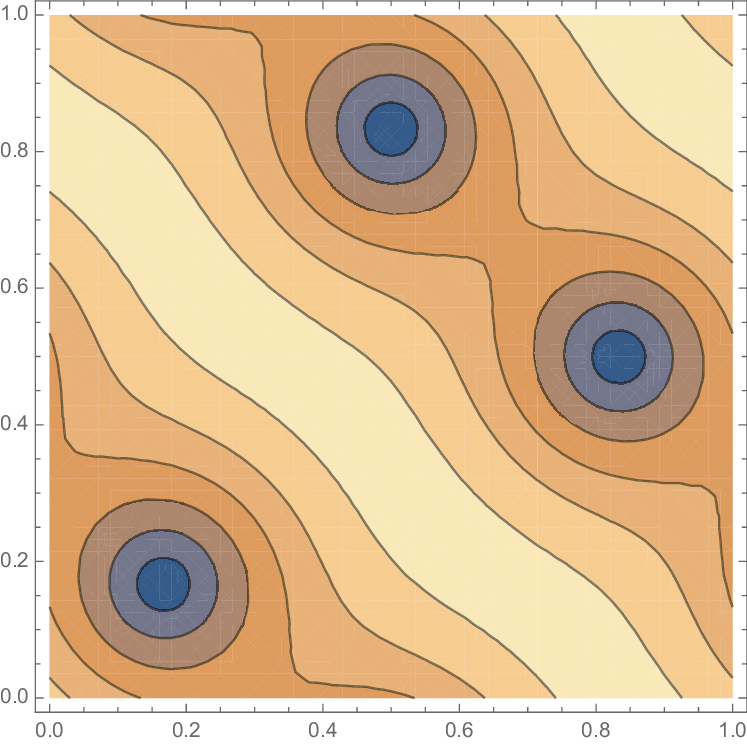} \\
{\small (e) $\langle \varphi^{(M=3)}_{{\rm{V}}}\rangle$}
\end{minipage}
\hfill
\begin{minipage}{0.32\textwidth}
\centering
\includegraphics[width=0.55\textwidth]{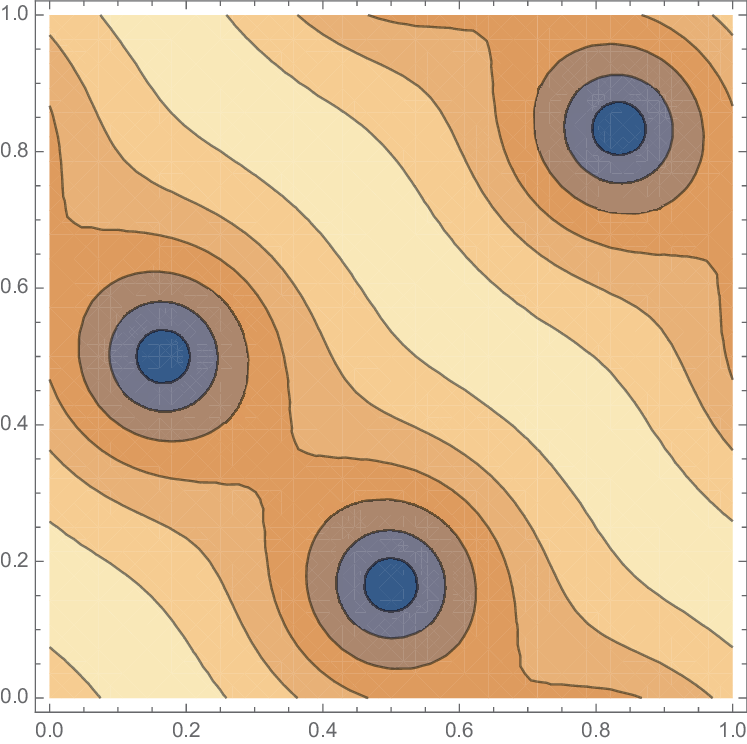} \\
{\small (f) $\langle \varphi^{(M=3)}_{{\rm{VI}}}\rangle$}
\end{minipage}
\caption{The zero points of the vacuum solution $\langle \varphi^{(M=3)}(z) \rangle$}
\label{fig_M_3}
\end{figure}
%

%
\paragraph{(i)$\,\varphi^{({\rm{I}})}_{\rm{vac}}(z)\\$}
%
We first consider the vacuum configuration
%
\begin{align}
\varphi^{({\rm{I}})}_{\rm{vac}}(z)\equiv
\langle \varphi_{\rm{I}}^{(M=3)}
(z)\rangle
&=a^{({\rm{I}})}\phi_0^{(0,M=3)}(z)+b^{({\rm{I}})}\phi_0^{(1,M=3)}(z)+c^{({\rm{I}})}\phi_0^{(2,M=3)}(z)
\label{sol1_{M=3}}
\end{align}
%
with numerical coefficients
%
\begin{align}
   a^{({\rm{I}})}=-0.0111561+0.0862216i  ,\quad
   b^{({\rm{I}})}=c^{({\rm{I}})}=0.0802481-0.0334494i.
   \label{Coe_M3_00_1}
\end{align}
%

%
This vacuum configuration $\varphi^{({\rm{I}})}_{\rm{vac}}(z)$ has three
zeros located at (see Figure \ref{fig_M_3}\,(a))
%
\begin{align}
z_1^{\rm{(I)}}=\frac{1+\tau}{6},\,\,
z_2^{\rm{(I)}}=\frac{1+\tau}{2},\,\,
z_3^{\rm{(I)}}=\frac{5+5\tau}{6}.
\end{align}
%
It is not difficult to see that the set of the zeros
$\{z_1^{\rm{(I)}},z_2^{\rm{(I)}},z_3^{\rm{(I)}}\}$ is invariant under the 
following symmetry transformations:
%
\begin{align}
H^{\rm{(I)}}
    &=\{ (t_{1}t_{\tau})^n(r_{\omega})^{2m},\,n=0,1,2\,({\rm{mod}} \,3),\,m=0,1\,
       ({\rm{mod}} \,2)\}
       \notag\\
    &= S_{3},
\end{align}
%
where $S_3$ denotes the symmetric group of degree $3$.

The above observation strongly suggests that the vacuum configuration 
$\varphi^{({\rm{I}})}_{\rm{vac}}(z)$ preserves the $H^{\rm{(I)}}$ symmetry. 
To verify this statement, let us investigate how the $H^{\rm{(I)}}$-invariance 
of the  $\varphi^{({\rm{I}})}_{\rm{vac}}(z)$ restricts the coefficients $a^{({\rm{I}})},b^{({\rm{I}})}$ and $c^{({\rm{I}})}$.
To this end, we impose the following relations on 
$\varphi^{({\rm{I}})}_{\rm{vac}}(z)$:
%
\begin{align}
    T_1 T_{\tau}(\varphi^{({\rm{I}})}_{\rm{vac}})(z) &\sim \varphi^{({\rm{I}})}_{\rm{vac}}(z),\label{T_1T_tau}\\
     (R_{\omega})^2(\varphi^{({\rm{I}})}_{\rm{vac}})(z) &\sim \varphi^{({\rm{I}})}_{\rm{vac}}(z),
     \label{R^2_00}
\end{align}
%
which are associated with $t_{1}t_{\tau}$ and $(r_{\omega})^{2}$ in $H^{(\rm{I})}$.

The requirement (\ref{T_1T_tau}) leads to the constraints on $a^{({\rm{I}})},b^{({\rm{I}})}$ and $c^{({\rm{I}})}$
%
\begin{align}
    \rho^2 a^{({\rm{I}})}=b^{({\rm{I}})}=c^{({\rm{I}})}\quad {\rm{or}} \quad
    a^{({\rm{I}})}=\rho^2b^{({\rm{I}})}=c^{({\rm{I}})}
    \quad {\rm{or}} \quad
    a^{({\rm{I}})}=b^{({\rm{I}})}=\rho^2c^{({\rm{I}})},
    \label{con_T_1T_tau}
\end{align}
%
where $\rho=e^{2\pi i/3}$.
On the other hand, Eq.\eqref{R^2_00} requires
%
\begin{align}
b^{({\rm{I}})}=c^{({\rm{I}})}.
\label{con_R^2_00}
\end{align}
%
Therefore, we find that if the vacuum configuration 
$\varphi^{({\rm{I}})}_{\rm{vac}}(z)$ holds the $H^{(\rm{I})}$-symmetry,
the coefficients $a^{({\rm{I}})},b^{({\rm{I}})}$ and $c^{({\rm{I}})}$
in Eq.\eqref{sol1_{M=3}} have to obey the relations
%
\begin{align}
    \rho^2 a^{({\rm{I}})}=b^{({\rm{I}})}=c^{({\rm{I}})}
    \qquad (\rho=e^{2\pi i/3}).
    \label{con_M3_00_1}
\end{align}
%
This is indeed the case.
Eq.\eqref{con_M3_00_1} is found to be consistent with the numerical result 
(\ref{Coe_M3_00_1}).
Thus, $\varphi^{({\rm{I}})}_{\rm{vac}}(z)$ preserves the $H^{(\rm{I})}$ symmetry
and the symmetry $G$ of the system is spontaneously broken to $H^{(\rm{I})}$.

%
\paragraph{(ii)$\,\varphi^{({\rm{II}})}_{\rm{vac}}(z)\\$}
%
The second vacuum configuration $\varphi^{({\rm{II}})}_{\rm{vac}}(z)$ is given by
%
\begin{align}
\varphi^{({\rm{II}})}_{\rm{vac}}(z)
 \equiv \langle \varphi_{\rm{II}}^{(M=3)}(z)\rangle
 =a^{({\rm{II}})}\phi_0^{(0,3)}(z)+b^{({\rm{II}})}\phi_0^{(1,3)}(z)+c^{({\rm{II}})}\phi_0^{(2,3)}(z),
 \label{sol2_{M=3}}
\end{align}
%
where the coefficients are numerically found as
%
\begin{align}
   a^{({\rm{II}})}=c^{({\rm{II}})}=0.0802481-0.0334494i ,\quad
   b^{({\rm{II}})}=-0.0111561+0.08622i.
   \label{Coe_M3_00_2}
\end{align}
%

%
The zeros of $\varphi^{({\rm{II}})}_{\rm{vac}}(z)$ 
are located at (see Figure \ref{fig_M_3}\,(b))
%
\begin{align}
z_1^{\rm{(II)}}=\frac{1+5\tau}{6},\,\,
z_2^{\rm{(II)}}=\frac{3+\tau}{6},\,\,
z_3^{\rm{(II)}}=\frac{5+3\tau}{6}.
\end{align}
%
The set of the zeros
$\{z_1^{\rm{(II)}},z_2^{\rm{(II)}},z_3^{\rm{(II)}}\}$ is invariant 
under the following symmetry transformations:
%
\begin{align}
H^{\rm{(II)}}
  &=\{ (t_{1}t_{\tau})^n(r_{\omega}^2t_1^{-1})^{m},\,n=0,1,2\,({\rm{mod}} \,3),
     \,m=0,1\,({\rm{mod}} \,2)\}
     \notag\\
  &= S_{3}.
\end{align}
%
Since the numerical coefficients in Eq.\eqref{sol2_{M=3}} satisfy
%
\begin{align}
a^{({\rm{II}})}= \rho^2 b^{({\rm{II}})}=c^{({\rm{II}})}
  \qquad (\rho = e^{2\pi i/3}),
    \label{con_M3_00_2}
\end{align}
%
the vacuum configuration $\varphi^{({\rm{II}})}_{\rm{vac}}(z)$ turns out
to satisfy the relations
%
\begin{align}
    T_1 T_{\tau}(\varphi^{({\rm{II}})}_{\rm{vac}})(z) &\sim \varphi^{({\rm{II}})}_{\rm{vac}}(z),\\
     (R_{\omega})^2 T_1^{-1}(\varphi^{({\rm{II}})}_{\rm{vac}})(z) &\sim \varphi^{({\rm{II}})}_{\rm{vac}}(z).
     \label{R^2T_1^-1_00}
\end{align}
%
This fact implies that the vacuum configuration 
$\varphi^{({\rm{II}})}_{\rm{vac}}(z)$
preserves the $H^{(\rm{II})}$ symmetry and the symmetry $G$ of the system
is spontaneously broken to $H^{(\rm{II})}$.

%
\paragraph{(iii)$\,\varphi^{({\rm{III}})}_{\rm{vac}}(z)\\$}
%
The third vacuum configuration $\varphi^{({\rm{III}})}_{\rm{vac}}(z)$ is given by
%
\begin{align}
\varphi^{({\rm{III}})}_{\rm{vac}}(z)\equiv
\langle \varphi_{\rm{III}}^{(M=3)}
(z)\rangle
&=a^{({\rm{III}})}\phi_0^{(0,3)}(z)+b^{({\rm{III}})}\phi_0^{(1,3)}(z)+c^{({\rm{III}})}\phi_0^{(2,3)}(z)
\label{sol3_{M=3}}
\end{align}
%
with
%
\begin{align}
   a^{({\rm{III}})}=b^{({\rm{III}})}=0.0802481-0.0334494i ,\quad
   c^{({\rm{III}})}=-0.0111561+0.08622i.
   \label{Coe_M3_00_3}
\end{align}
%

%
The zeros of $\varphi^{({\rm{III}})}_{\rm{vac}}(z)$
are located at (see Figure \ref{fig_M_3}\,(c)) 
%
\begin{align}
z_1^{\rm{(III)}}=\frac{1+3\tau}{6},\,\,
z_2^{\rm{(III)}}=\frac{3+5\tau}{6},\,\,
z_3^{\rm{(III)}}=\frac{5+\tau}{6}.
\end{align}
%
The set of the zeros
$\{z_1^{\rm{(III)}},z_2^{\rm{(III)}},z_3^{\rm{(III)}}\}$ is invariant 
under the following symmetry transformations:
%
\begin{align}
H^{\rm{(III)}}
  &=\{ (t_{1}t_{\tau})^n(r_{\omega}^2t_1)^{m},\,n=0,1,2\,({\rm{mod}} \,3),
     \,m=0,1\,({\rm{mod}} \,2)\}
     \notag\\
  &=S_3. 
\end{align}
%
Since the numerical coefficients in Eq.\eqref{sol3_{M=3}} satisfy
%
\begin{align}
a^{({\rm{III}})}=  b^{({\rm{III}})}=\rho^2 c^{({\rm{III}})}
  \qquad (\rho=e^{2\pi i/3}),
    \label{con_M3_00_3}
\end{align}
%
we find that the vacuum configuration $\varphi^{({\rm{III}})}_{\rm{vac}}(z)$
obeys the relations
%
\begin{align}
    T_1 T_{\tau}(\varphi^{({\rm{III}})}_{\rm{vac}})(z) &\sim \varphi^{({\rm{III}})}_{\rm{vac}}(z),\\
     (R_{\omega})^2 T_1(\varphi^{({\rm{III}})}_{\rm{vac}})(z) &\sim \varphi^{({\rm{III}})}_{\rm{vac}}(z).
     \label{R^2T_1_00}
\end{align}
%
This fact implies that the vacuum configuration 
$\varphi^{({\rm{III}})}_{\rm{vac}}(z)$ preserves the $H^{(\rm{III})}$-symmetry 
and the symmetry $G$ of the system is spontaneously broken to $H^{(\rm{III})}$.

%
\paragraph{(iv)$\,\varphi^{({\rm{IV}})}_{\rm{vac}}(z)\\$}
%
The forth vacuum configuration $\varphi^{({\rm{IV}})}_{\rm{vac}}(z)$ is given by
%
\begin{align}
\varphi^{({\rm{IV}})}_{\rm{vac}}(z)\equiv
\langle \varphi_{\rm{IV}}^{(M=3)}
(z)\rangle
&=a^{({\rm{IV}})}\phi_0^{(0,3)}(z)+b^{({\rm{IV}})}\phi_0^{(1,3)}(z)+c^{({\rm{IV}})}\phi_0^{(2,3)}(z)
\label{sol5_{M=3}}
\end{align}
%
with
%
\begin{align}
   a^{({\rm{IV}})}=-0.0160683-0.0854426i,\quad
   b^{({\rm{IV}})}=c^{({\rm{IV}})}=0.0820296+0.0288056i.
   \label{Coe_M3_00_4}
\end{align}
%

%
The zeros of $\varphi^{({\rm{IV}})}_{\rm{vac}}(z)$
are located at (see Figure \ref{fig_M_3}\,(d))
%
\begin{align}
z_1^{\rm{(IV)}}=\frac{1+5\tau}{6},\,\,
z_2^{\rm{(IV)}}=\frac{1+\tau}{2},\,\,
z_3^{\rm{(IV)}}=\frac{5+\tau}{6}.
\end{align}
%
The set of $\{z_1^{\rm{(IV)}},z_2^{\rm{(IV)}},z_3^{\rm{(IV)}}\}$ 
is invariant under the following symmetry transformations:
%
\begin{align}
H^{\rm{(IV)}}
  &=\{ (t_{1}^{-1}t_{\tau})^n(r_{\omega})^{2m},\,n=0,1,2\,({\rm{mod}} \,3),
    \,m=0,1\,({\rm{mod}} \,2)\}
    \notag\\
  &= S_{3}.
\end{align}
%
Since the numerical coefficients in Eq.\eqref{sol5_{M=3}} satisfy
%
\begin{align}
    \rho a^{({\rm{IV}})}=  b^{({\rm{IV}})}= c^{({\rm{IV}})}
    \qquad (\rho=e^{2\pi i/3}),
    \label{con_M3_00_4}
\end{align}
%
we find that the vacuum configuration $\varphi^{({\rm{IV}})}_{\rm{vac}}(z)$
obeys the relations
%
\begin{align}
    T_1^{-1} T_{\tau}(\varphi^{({\rm{IV}})}_{\rm{vac}})(z) &\sim \varphi^{({\rm{IV}})}_{\rm{vac}}(z),\label{T_1^-1T_tau}\\
     (R_{\omega})^2 (\varphi^{({\rm{IV}})}_{\rm{vac}})(z) &\sim \varphi^{({\rm{IV}})}_{\rm{vac}}(z).
\end{align}
%
This fact implies that the vacuum configuration $\varphi^{({\rm{IV}})}_{\rm{vac}}(z)$
preserves the $H^{(\rm{IV})}$-symmetry and the symmetry $G$ of the system
is spontaneously broken to $H^{(\rm{IV})}$.

%
\paragraph{(v)$\,\varphi^{({\rm{V}})}_{\rm{vac}}(z)\\$}
%
The fifth vacuum configuration $\varphi^{({\rm{V}})}_{\rm{vac}}(z)$ is given by
%
\begin{align}
\varphi^{({\rm{V}})}_{\rm{vac}}(z)\equiv
\langle \varphi_{\rm{V}}^{(M=3)}
(z)\rangle
&=a^{({\rm{V}})}\phi_0^{(0,3)}(z)+b^{({\rm{V}})}\phi_0^{(1,3)}(z)+c^{({\rm{V}})}\phi_0^{(2,3)}(z)
\label{sol4_{M=3}}
\end{align}
%
with
%
\begin{align}
a^{({\rm{V}})}=b^{({\rm{V}})}=0.0820296+0.0288056i,\quad 
   c^{({\rm{V}})}=-0.0160683-0.0854426i.
   \label{Coe_M3_00_5}
\end{align}
%

%
The zeros of $\varphi^{({\rm{V}})}_{\rm{vac}}(z)$ 
are located at (see Figure \ref{fig_M_3}\,(e))
%
\begin{align}
z_1^{\rm{(V)}}=\frac{1+\tau}{6},\,\,
z_2^{\rm{(V)}}=\frac{3+5\tau}{6},\,\,
z_3^{\rm{(V)}}=\frac{5+3\tau}{6}.
\end{align}
%
The set of the zeros
$\{z_1^{\rm{(V)}},z_2^{\rm{(V)}},z_3^{\rm{(V)}}\}$ is invariant 
under the following symmetry transformations:
%
\begin{align}
H^{\rm{(V)}}
  &=\{ (t_{1}^{-1}t_{\tau})^n(r_{\omega}^2 t_1^{-1})^{m},\,n=0,1,2\,({\rm{mod}} \,3),
     \,m=0,1\,({\rm{mod}} \,2)\}
     \notag\\
  &= S_{3}.
\end{align}
%
Since the numerical coefficients in Eq.\eqref{sol4_{M=3}} satisfy
%
\begin{align}
     a^{({\rm{V}})}=  b^{({\rm{V}})}= \rho c^{({\rm{V}})}
     \qquad (\rho=e^{2\pi i/3}),
    \label{con_M3_00_5}
\end{align}
%
we find that the vacuum configuration $\varphi^{({\rm{V}})}_{\rm{vac}}(z)$ 
obeys the relations
%
\begin{align}
    T_1^{-1} T_{\tau}(\varphi^{({\rm{V}})}_{\rm{vac}})(z) &\sim \varphi^{({\rm{V}})}_{\rm{vac}}(z),\label{T_1^-1T_tau_}\\
     (R_{\omega})^2 T_1^{-1}(\varphi^{({\rm{V}})}_{\rm{vac}})(z) &\sim \varphi^{({\rm{V}})}_{\rm{vac}}(z).
\end{align}
%
This fact implies that the vacuum configuration $\varphi^{({\rm{V}})}_{\rm{vac}}(z)$ 
preserves the $H^{(\rm{V})}$-symmetry and the symmetry $G$ of the system
is spontaneously broken to $H^{(\rm{V})}$.

%
\paragraph{(vi)$\,\varphi^{({\rm{VI}})}_{\rm{vac}}(z)\\$}
%
The sixth vacuum configuration $\varphi^{({\rm{VI}})}_{\rm{vac}}(z)$ is given by
%
\begin{align}
\varphi^{({\rm{VI}})}_{\rm{vac}}(z)\equiv
\langle \varphi_{\rm{VI}}^{(M=3)}
(z)\rangle
&=a^{({\rm{VI}})}\phi_0^{(0,3)}(z)+b^{({\rm{VI}})}\phi_0^{(1,3)}(z)+c^{({\rm{VI}})}\phi_0^{(2,3)}(z)
\label{sol6_{M=3}}
\end{align}
%
with
%
\begin{align}
a^{({\rm{VI}})}=c^{({\rm{VI}})}=0.0820296+0.0288056i,\quad 
   b^{({\rm{V}})}=-0.0160683-0.0854426i.
   \label{Coe_M3_00_6}
\end{align}
%

%
The zeros of $\varphi^{({\rm{VI}})}_{\rm{vac}}(z)$
are located at (see Figure \ref{fig_M_3}\,(f))
%
\begin{align}
z_1^{\rm{(VI)}}=\frac{1+3\tau}{6},\,\,
z_2^{\rm{(VI)}}=\frac{3+\tau}{6},\,\,
z_3^{\rm{(VI)}}=\frac{5+5\tau}{6}.
\end{align}
%
The set of $\{z_1^{\rm{(VI)}},z_2^{\rm{(VI)}},z_3^{\rm{(VI)}}\}$ 
is invariant under the following symmetry transformations:
%
\begin{align}
H^{\rm{(VI)}}
   &=\{ (t_{1}^{-1}t_{\tau})^n(r_{\omega}^2 t_1)^{m},\,n=0,1,2\,({\rm{mod}} \,3),
     \,m=0,1\,({\rm{mod}} \,2)\}
     \notag\\
   &= S_{3}.
\end{align}
%
Since the numerical coefficients in Eq.\eqref{sol6_{M=3}} satisfy
%
\begin{align}
     a^{({\rm{VI}})}= \rho  b^{({\rm{VI}})}= c^{({\rm{VI}})}
     \qquad (\rho=e^{2\pi i/3}),
    \label{con_M3_00_6}
\end{align}
%
we find that the vacuum configuration $\varphi^{({\rm{VI}})}_{\rm{vac}}(z)$
obeys the relations
%
\begin{align}
    T_1^{-1} T_{\tau}(\varphi^{({\rm{VI}})}_{\rm{vac}})(z) &\sim \varphi^{({\rm{VI}})}_{\rm{vac}}(z),\\
     (R_{\omega})^2 T_1(\varphi^{({\rm{VI}})}_{\rm{vac}})(z) &\sim \varphi^{({\rm{VI}})}_{\rm{vac}}(z).
\end{align}
%
This fact implies that the vacuum configuration $\varphi^{({\rm{VI}})}_{\rm{vac}}(z)$
preserves the $H^{(\rm{VI})}$-symmetry and the symmetry $G$ of the system
is spontaneously broken to $H^{(\rm{VI})}$.

It follows from the above analysis that the symmetry $G$ of the system
is spontaneously broken to 
%
\begin{align}
    G\,\,\to\,\, H^{({\rm{J}})} \qquad (\rm{J}=I,II,\cdots,VI)
\end{align}
%
for the vacuum configuration $\varphi^{({\rm{J}})}_{\rm{vac}}(z)$.
This fact explains the reason why there appear the six degenerate vacuum
configurations to minimize the potential.
The symmetry group $G$ can be decomposed into the right coset
with respect to $H^{(\rm{I})}$ as
%
\begin{align}
    G=H^{({\rm{I}})} \cup (t_1 H^{({\rm{I}})})
    \cup (t_1^2 H^{({\rm{I}})}) \cup(r_{\omega} H^{({\rm{I}})})
    \cup (t_1 r_{\omega} H^{({\rm{I}})})
    \cup (t_1^2 r_{\omega} H^{({\rm{I}})}).
\end{align}
%
This result indicates that there should exist six degenerate configurations as
the vacua.
In fact, the configurations 
$\varphi^{({\rm{II}})}_{\rm{vac}},\varphi^{({\rm{III}})}_{\rm{vac}},\cdots,\varphi^{({\rm{VI}})}_{\rm{vac}}$
can be obtained from $\varphi^{({\rm{I}})}_{\rm{vac}}$ as
%
\begin{align}
\varphi^{({\rm{I}})}_{\rm{vac}}
 \ \ \stackrel{X}{\longrightarrow}\ \ 
  \begin{cases}
    \varphi^{({\rm{II}})}_{\rm{vac}} & \textrm{for}\ X=(T_{1})^{2},\\
    \varphi^{({\rm{III}})}_{\rm{vac}} & \textrm{for}\ X=T_{1},\\  
    \varphi^{({\rm{IV}})}_{\rm{vac}} & \textrm{for}\ X=R_{\omega},\\
    \varphi^{({\rm{V}})}_{\rm{vac}} & \textrm{for}\ X=(T_{1})^{2}R_{\omega},\\
    \varphi^{({\rm{VI}})}_{\rm{vac}} & \textrm{for}\ X=T_{1}R_{\omega}.
  \end{cases}
\end{align}
%
Therefore, the six vacuum configurations are connected each other
by the broken symmetry transformations of the coset $G/H$, and hence they are
physically equivalent.

%
\section{Stability of the vacuum}
\label{Stability}
%
In the previous section, the vacuum configurations were obtained approximately
by restricting the scalar field to a linear combination of the
lowest-mode eigenfunctions $\phi_{0}^{(j,M)}(z)\ (j=0,1,\cdots,M-1)$
on the magnetized torus.
In this section, we demonstrate that such vacuum configurations are
perturbatively stable, provided the condition \eqref{area_condition} 
is satisfied.

Let $\varphi^{(0)}(z) = \sum_{j=0}^{M-1}a_{0}^{(j)} \phi_{0}^{(j,M)}(z)$
be a solution that minimizes the potential $\widetilde{V}[\varphi^{(0)}]$
within the lowest-mode approximation.
We consider a small fluctuation around $\varphi^{(0)}(z)$ as
%
\begin{align}
    \varphi(z)=\varphi^{(0)}(z)+\delta\varphi(z),
    \label{pert}
\end{align}
%
where $\delta\varphi(z)$ represents a perturbation and is expanded as
%
\begin{align}
\delta\varphi(z) = \sum_{n=1}^{\infty}\sum_{j=0}^{M-1}a_{n}^{(j)} 
                   \phi_{n}^{(j,M)}(z).
\label{deltaphi}
\end{align}
%

%
By substituting Eq.\eqref{pert} into Eq.\eqref{H_scalar_zerodim}, we obtain
the expansion
%
\begin{align}
    \widetilde{V}[\varphi^{(0)}+\delta\varphi]
    =\widetilde{V}[\varphi^{(0)}]
      +\frac{\delta \widetilde{V}}{\delta \varphi}[\varphi^{(0)}]\cdot \delta\varphi
      +\frac{1}{2}\frac{\delta^2 \widetilde{V}}{\delta \varphi^2}[\varphi^{(0)}]\cdot 
        (\delta\varphi)^2 +\cdots.
    \label{stability}
\end{align}
%
Since $\varphi^{(0)}(z)$ is supposed to be a solution to minimize 
$\widetilde{V}[\varphi^{(0)}]$, $\varphi^{(0)}(z)$ is a stationary point of
$\widetilde{V}[\varphi^{(0)}]$, i.e. 
$\frac{\delta \widetilde{V}}{\delta\varphi}[\varphi^{(0)}] = 0$.
Hence, the the second term on the right-hand side of
Eq.\eqref{stability} vanishes.

The second-order variation is explicitly given by
%
\begin{align}
    &\frac{1}{2}\frac{\delta^2 \widetilde{V}}{\delta \varphi^2}[\varphi^{(0)}]\cdot ( \delta\varphi)^2 
    \notag \\
    &=\int_{T^2}d^2 z \, \left\{\delta\varphi^{\ast}
       \left[-2(D_{z}D_{\bar{z}}+D_{\bar{z}}D_{z})-{\mu^{\prime}}^2\right]
        \delta\varphi
        + 2\lambda'|\varphi^{(0)}|^2|\delta\varphi|^{2}
        +\lambda^{\prime}(\varphi^{(0)}\delta\varphi^{\ast}
          +\varphi^{(0)\ast}\delta\varphi)^2\right\}.
\label{2th}
\end{align}
%
When the condition \eqref{area_condition} is satisfied, the first term
of Eq.\eqref{2th} is positive-definite for any nonvanishing fluctuations
of the form \eqref{deltaphi}.
Consequently, the second-order variation \eqref{2th} is strictly positive
for $\delta\varphi \ne 0$.
Therefore, we conclude that the configuration $\varphi^{(0)}(z)$ is
a local minimum of the functional $\widetilde{V}$ and is perturbatively stable
as long as the condition \eqref{area_condition} is satisfied.

The above argument indicates that when the area $A_{T^{2}}$
of the torus satisfies the inequality
%
\begin{align}
\frac{4\pi M}{\mu^{2}}\Big(K+\frac{1}{2}\Big)
  < A_{T^2} < \frac{4\pi M}{\mu^{2}}\Big(K+\frac{3}{2}\Big),
\label{area_condition2}
\end{align}
%
for some non-negative integer $K$, a stable vacuum configuration 
requires the inclusion of modes up to level $K$, namely
%
\begin{align}
\varphi^{(0)}(z) = \sum_{n=0}^{K}\sum_{j=0}^{M-1}a_{n}^{(j)} 
                   \phi_{n}^{(j,M)}(z).
\label{deltaphi_K}
\end{align}
%

%
\section{Conclusions and discussions}
\label{sec:final}
%
In this paper, we have investigated the vacuum structure of a complex
scalar field on a magnetized two-dimensional torus.
We have shown that the system exhibits a critical area, above which
the vacuum expectation value becomes nonvanishing.
In contrast to conventional Higgs mechanisms, the nonvanishing vacuum
expectation value necessarily depends on the coordinate of the torus.

In Section \ref{sec:VEV}, we have obtained one, two and six 
vacuum configurations to minimize the potential for $M=1,2$ and $3$,
respectively, within the lowest-mode approximation.
In Section \ref{sec:symmetry}, we have discussed the symmetries of the vacuum configurations.
We have first shown that the system has
$G=(\mathbb{Z}_M^1 \times \mathbb{Z}_M^{\tau} )\rtimes {\mathbb{Z}}_4^{\omega}$
discrete symmetry, where $\mathbb{Z}_M^1 \times \mathbb{Z}_M^{\tau}$ is 
the discrete translational symmetry and ${\mathbb{Z}}_4^{\omega}$ is the $\pi/2$
rotational one for the torus moduli $\tau=i$.
We have next examined the symmetries of the vacuum configurations
for $M=1,2$ and $3$.
For $M=1$, the symmetry of the system is $G=\mathbb{Z}_{4}^{\omega}$
and the symmetry is preserved by the vacuum.
For $M=2$, there are two degenerate vacuum configurations, and 
they are found to break the symmetry 
$G=(\mathbb{Z}_{M=2}^1 \times \mathbb{Z}_{M=2}^{\tau} )\rtimes 
{\mathbb{Z}}_4^{\omega}$
of the system to $H=\mathbb{Z}_{2}^{\prime} \times \mathbb{Z}_{4}^{\prime}$.
For $M=3$, there are six degenerate vacuum configurations, and the symmetry
of the system is spontaneously broken from 
$G=(\mathbb{Z}_{M=3}^1 \times \mathbb{Z}_{M=3}^{\tau} )\rtimes 
{\mathbb{Z}}_4^{\omega}$ to $H=S_{3}$.
The six vacuum configurations are connected to each other by the 
transformations belonging to $G/H$.
Finally, in Section \ref{Stability}, we have discussed the stability
of the vacuum configurations obtained in the lowest-mode approximation,
and shown that they are perturbatively stable.

In this work, we have focused on the specific case of $\tau=i$ and $M=1,2$ and $3$.
It would be of interest to extend the analysis to general values of $\tau$
and $M$, because the vacuum structure may depend sensitively on these
parameters.
Furthermore, our work has been limited to the lowest-mode approximation.
The analysis beyond the approximation should be performed.
Those extensions will be left for future study.

Unlike conventional scenarios where symmetry breaking is induced by
constant vacuum expectation values, the present model necessarily
involves coordinate-dependent vacua.
This feature may lead to distinctive phenomenological consequences,
for instance in the mass spectra of gauge bosons and fermions in
extra-dimensional models.

Finally, we comment on a possible connection to condensed matter systems.
Wavefunctions on a magnetized torus with flux $M$ possess $|M|$ zeros,
which may be interpreted as vortices.
Due to the compactness of the torus, finding the configuration of such vortices
is highly nontrivial.
Understanding the vacuum structure from this perspective may provide
an interesting avenue for future research.

%
\section*{Acknowledgements}
%
This work was partially supported by Japan Society for the Promotion of Science (JSPS) KAKENHI Grant No. 23K03416(M.S.).
M.T. was in part supported by the Sumitomo Foundation under Grant for Basic Science Research (Grant No. 2502479).
H.I. was supported by the Fuujukai Foundation.
This work was also partly supported by MEXT Promotion of Distinctive Joint
Research Center Program JPMXP0723833165 and Osaka Metropolitan
University Strategic Research Promotion Project (Development of
International Research Hubs).

\bibliographystyle{utphys}
\bibliography{main}
\end{document}